\documentclass[useAMS,usenatbib]{mn2e}
\usepackage{amsmath}
\usepackage{amssymb}
\usepackage{multirow}
\usepackage{graphicx}
\usepackage{epstopdf}
\usepackage{color}

\newcommand{\Tb}{T_{\rm{b}}}
\newcommand{\Ts}{T_{\rm s}}
\newcommand{\Tcmb}{T_{\rm CMB}}
\newcommand{\Tk}{T_{\rm k}}
\newcommand{\Tc}{T_{\rm c}}
\newcommand{\xHI}{x_{\rm HI}}

\title[Quantiles and 21-cm Non-Gaussianity] {Quantiles as Robust
  Probes of Non-Gaussianity in 21-cm Images} \author[Banet et
  al.]{Alon Banet$^{1}$\thanks{E-mail: alon.banet@gmail.com}, Rennan
  Barkana$^{1}$, Anastasia Fialkov$^{2}$, Or Guttman$^{1}$, \\ $^{1}$
  School of Physics and Astronomy, Tel Aviv University, Tel Aviv
  69978, Israel\\ $^{2}$ Institute of Astronomy, University of
  Cambridge, Madingley Road, Cambridge CB3 0HA, UK}

\begin{document}
\pagerange{\pageref{firstpage}--\pageref{lastpage}} \pubyear{2020}
\maketitle

\label{firstpage}

\begin{abstract}
The early epoch in which the first stars and galaxies formed is among
the most exciting unexplored eras of the Universe. A major research
effort focuses on probing this era with the 21-cm spectral line of
hydrogen. While most research focused on statistics like the 21-cm
power spectrum or the sky-averaged global signal, there are other ways
to analyze tomographic 21-cm maps, which may lead to novel
insights. We suggest statistics based on quantiles as a method to
probe non-Gaussianities of the 21-cm signal. We show that they can be
used in particular to probe the variance, skewness, and kurtosis of
the temperature distribution, but are more flexible and robust than
these standard statistics. We test these statistics on a range of
possible astrophysical models, including different galactic halo
masses, star-formation efficiencies, and spectra of the X-ray heating
sources, plus an exotic model with an excess early radio background.
Simulating data with angular resolution and thermal noise as expected
for the Square Kilometre Array (SKA), we conclude that these
statistics can be measured out to redshifts above 20 and offer a
promising statistical method for probing early cosmic history.
\end{abstract}

\begin{keywords}
dark ages, reionization, first stars -- cosmology: theory -- galaxies:
high redshift
\end{keywords}

\section{Introduction}
\label{Sec:Intro}

Ever since Penzias and Wilson discovered the Cosmic Microwave
Background (CMB) in 1964, cosmologists have had a good understanding
of the Universe in its early stages.\ Meanwhile, modern telescopes
have allowed astronomers to study astronomical objects in the more
recent Universe, reaching times as early as a billion years after the
Big Bang. Despite tremendous progress in recent decades, the exciting
period in between, in which stars and galaxies first formed and
evolved, remains largely unobserved to this day.

While a few bright galaxies that date back to 400~Myr after the Big
Bang have been detected directly via telescopes, it is thought that
most of the early stars are distributed in a large number of very
small galaxies, making them difficult to observe directly. The most
promising probe of these early times is the spin-flip transition of
neutral hydrogen. Since redshift acts as the line-of-sight dimension,
it can be used to produce a 3-D tomography map of the cosmic gas.

The brightness temperature of the 21-cm signal (which is measured
relative to the CMB temperature) is determined by the spin
temperature, which denotes the abundance of the excited level of the
hyperfine split of hydrogen relative to the ground level. The spin
temperature is affected by astrophysical and cosmological events, so
therefore it may allow us to study star and galaxy formation within
dark matter halos, as well as phenomena like cosmic reionization and
early cosmic heating.

Previous studies have shown that the 21-cm signal should have large
spatial fluctuations, which stem not only from reionization at low
redshifts, but also from fluctuations in the Ly$\alpha$ intensity
during the Ly$\alpha$ coupling era \citep{Barkana:2005} and
fluctuations in the X-ray background during the era of cosmic heating
\citep{Pritchard:2007}. Research over this past decade has focused on
the statistics of these fluctuations and in particular, on the 21-cm
power spectrum, a highly promising measure of the 21-cm signal. Since
the foregrounds are expected to have a smooth spectrum, the power
spectrum may be measured in the near future. Another approach, pursued
by both theorists and experimentalists is the sky-averaged global
21-cm signal, which could prove useful in independently constraining
the parameters of the early universe. The first claimed detection of a
cosmological 21-cm signal is of the global signal at cosmic dawn, by
the EDGES experiment \citep{Bowman:2018}. The surprisingly-deep
absorption, if confirmed, would require an exotic explanation, such as
an interaction with dark matter that cools the baryons
\citep{Barkana:2018}, or an enhanced early radio background (discussed
further below).

These methods span only part of the richness the 21-cm signal holds.
In particular, they are not able to probe non-Gaussianities in the
signal caused by the non-linear processes described above. In the near
future the Square Kilometre Array (SKA) should provide a detailed 3-D
map of the 21-cm fluctuation signal. However, the topic of analyzing
such images has received only limited attention
\citep{Koopmans:2015}. The aim of this study is to explore new ways of
studying these maps, in order to gain new insights about the 21-cm
signal. In the near future, maps will likely have a fairly low
signal-to-noise ratio, so use of averaging through statistics will be
necessary, but it is possible to go beyond the power spectrum. Use of
the 21-cm bispectrum has been explored
\citep[e.g.,][]{Bharadwaj,Majumdar,Trott}. A number of papers have
explored use of the probability distribution function of 21-cm
brightness temperature
\citep[e.g.,][]{Ciardi,Fur04,Mellema,Ichikawa,Mondal}, and in
particular the skewness and kurtosis statistics, mostly during the
reionization era \citep{Wyithe,LOFAR,Watkinson,Kubota,Kitt} and out to
cosmic dawn \citep{Shimabukuro,WatkinsonCD}. We explore these standard
statistics within a wide range of possible models, and also suggest
new statistics that serve as a more robust and flexible measure of
non-Gaussian characteristics and can help us explore the evolution of
the signal and understand the processes affecting it.

This paper is structured as follows. In section~2 we present the
details of how we simulated 21-cm images, laying out the assumed
models and their main parameters (2.1), and how we imitated
observational aspects corresponding to resolution and noise (2.2).  In
section~3 we present our statistical methods and results, laying out
measures based on quantiles (3.1), finding average radial profiles
(3.2), showing how we corrected for thermal noise (3.3), plotting the
variance and our alternative (the quantile average) (3.4), exploring
the extra flexibility of the quantile average (3.5), plotting the
skewness and our alternative (the quantile difference) (3.6), as well
as the kurtosis and our alternative (the normalized quantile average)
(3.7). Finally, we summarize and conclude in section~4.

\section{Simulated 21-cm tomography maps}
\label{Sec:Simulation}

We obtain the 21-cm image boxes using a semi-numerical simulation
\citep[e.g.,][]{21cmfast} in a box that is 384~Mpc on a side, with
3~Mpc resolution (comoving units), as described by
\citet{Cohen:2017}. The observed brightness temperature (relative to
the CMB) depends on the spin temperature $\Ts$, the neutral fraction
$\xHI$, and the baryonic overdensity $\delta$, as follows
\citep[e.g.,][]{Madau,Barkana:2016}:

\begin{equation}
\label{Eq:Tb}  
\Tb \propto \xHI(1+\delta) \left(1-\frac{\Tcmb}{\Ts}\right)\ .
\end{equation}
The spin temperature plays an important role in the evolution of the
signal, and $\Ts$ can be expressed as a weighted mean
\citep{Barkana:2016}:
\begin{equation}
\label{Eq:Ts} 
\Ts^{-1}=\frac{\Tcmb^{-1}+x_{\rm c}\Tk^{-1}+x_{\rm \alpha} \Tc^{-1}}{1+x_{\rm c}+x_{\rm \alpha}}\ ,
\end{equation}
where $\Tk$ is the kinetic temperature of the gas, $\Tc$ is
        the effective (color) Ly$\alpha$ temperature (which is very
        close to $\Tk$), and $x_{\rm c}$, $x_{\rm\alpha}$ are the
        coupling coefficients for collisions and Ly$\alpha$
        scattering, respectively.

At redshifts above $z \sim 200$, $\Tk$ was close to $\Tcmb$, causing
the signal to vanish. As the universe expanded the gas cooled
adiabatically, faster than the CMB, while atomic collisions kept the
spin temperature coupled to $\Tk$, leading to an absorption
signal. Eventually, the gas density decreased enough to make
collisional coupling ineffective, the radiative coupling of $\Ts$ to
$\Tcmb$ dominated and the signal diminished. As star formation began,
Ly$\alpha$ photons were emitted and coupled $\Ts$ to $\Tk$ via the
Wouthuysen-Field \citep{Wouthuysen:1952, Field:1958}
effect. Meanwhile, X-ray sources started heating the cosmic gas and UV
photons ionized the gas around galaxies, creating ionized bubbles and
initiating the process of cosmic reionization. It is useful to define
three milestone redshifts. A typical theoretical set of definitions
would be: Ly$\alpha$ coupling, defined as when the mean $x_{\rm
  \alpha}=1$; the heating transition, defined as when the mean
$\Tk=\Tcmb$; and the mid-point of reionization, at which the mean
$\xHI=0.5$. However, in plots below we adopt a modified set of
milestone redshifts, defined phenomenologically using peak redshifts
of our main measure of the signal (the quantile average, discussed
below in section~\ref{s:qave} and shown for our various models in
Figure~\ref{fig:AveAll}).

\subsection{Models and parameters}

To illustrate our method of exploring the characteristics of 21-cm
intensity maps, we used several models that differ in their input
astrophysical parameters. Given the early state of 21-cm observations,
the details of astrophysics at high redshift are still highly
uncertain, and it is important to consider a wide range of possible
models. The following are the main parameters of our models
\citep{Cohen:2017}:

\begin{enumerate}
\item Star formation efficiency (SFE) - the fraction of gas that is
  converted into stars, out of the gas that falls into star-forming
  dark matter halos. The overall SFE depends on the details of the
  process of star formation as well as the dominant feedback
  mechanisms. It strongly affects the 21-cm signal by influencing the
  amount of radiation produced by stars. For otherwise identical
  astrophysical parameters, a higher SFE implies an earlier onset of
  Ly$\alpha$ coupling, and a faster build-up of X-ray and ionizing
  radiation backgrounds; hence, a high SFE value shifts the
  cosmological 21-cm signal milestones to higher redshifts.
\item Cooling mass - the minimum halo mass in which there is
  significant gas cooling (and thus star formation). It depends on the
  cooling channels of the gas in halos, and is best described in terms
  of a minimum circular velocity $V_{\rm c}$. In atomic cooling halos,
  stars form with masses down to the cooling threshold of atomic
  hydrogen, given by $V_{\rm c}=16.5$~km~s$^{-1}$. As an example of
  strong feedback, we consider a model of ``Massive'' halos in which
  stars only form in halos with masses of at least 100 times the mass
  required for atomic cooling, which corresponds to $V_{\rm
    c}=76.5$~km~s$^{-1}$. In this model star formation is delayed, so
  that the 21-cm milestones are shifted to lower redshift values.
\item The spectrum of early X-ray sources. The mean free path of an
  X-ray photon is proportional to $E_{\rm photon}^3$, and thus soft
  X-rays have relatively short mean free paths and therefore they are
  absorbed soon after emission, heating the local gas before suffering
  significant energy loss due to redshift effects. Thus, soft X-ray
  sources cause large spatial fluctuations in the gas temperature
  during cosmic heating. However, the most plausible sources of cosmic
  heating are X-ray binaries, which are expected to have a relatively
  hard spectrum \citep{Mirabel:2011,Fragos}. Due to their long mean
  free path, the photons emitted from such sources will be absorbed
  late, after having lost a significant part of their energy as a
  result of cosmological redshift \citep{Fialkov:2014b}. Hence, a hard
  X-ray spectrum leads to cosmic heating at a later time and reduces
  the fluctuations in $\Tk$. Our standard assumption is a hard X-ray
  spectrum, but given the current uncertainty in the properties of
  high-redshift sources, we also consider a model with a soft X-ray
  spectrum.
\item X-ray radiation efficiency - proportional to the ratio of the
  X-ray luminosity to the star formation rate (SFR). It is normalized
  so that unity corresponds to low metallicity, low redshift starburst
  galaxies \citep{Mineo:2012}. Higher X-ray efficiency leads to
  earlier cosmic heating.
\item Excess radio background radiation. In order to explain the EDGES
  measurement of the global 21-cm signal at $z=17.2$ (which
  corresponds to $\nu=78.2 \rm{MHz}$) \citep{Bowman:2018}, we consider
  an example of an exotic model with a greatly enhanced early radio
  background \citep{Bowman:2018,Feng:2018,Fialkov:2019}. In this model
  the background temperature at redshift $z$ is modified to:

\begin{equation}
\label{Eq:Trad}  
T_{\rm rad}=\Tcmb(1+z)\left[1+A_{\rm r}\left(\frac{\nu_{\rm obs}}{\rm{78~MHz}}\right)^\beta\right] ,
\end{equation}

where $\nu_{\rm obs}$ is the observed frequency,
                $A_{\rm r}$ is the amplitude defined relative to the
                CMB temperature, and $\beta=-2.6$ is the spectral
                index, assumed to follow the shape of the observed
                radio background. The radio background enhances the
                21-cm signal when there is absorption, i.e., when $\Ts
                \ll T_{\rm rad}$.§

\end{enumerate}

For our study we chose four models from \citealp{Cohen:2017}
        plus an exotic model with an excess radio background, chosen
        to be generally consistent with the EDGES measurement
        \citep{Fialkov:2019}. The full parameters are listed in
        Table~\ref{table:casesparam}.

\begin{table}
\centering
\begin{tabular}{ c l l c c} 

\hline
\textbf{Model}    & \textbf{$f_*$} & \textbf{$f_X$} & \textbf{SED} & \textbf{Halo type}\\
\hline
Standard& 0.05  & 1   & Hard & Atomic cooling \\
$\rm\#$53&       &     &      & ($V_{\rm c}=16.5$ km/s) \\
\hline
Low-Efficiency& 0.005 & 0.1 & Hard & Atomic cooling \\
$\rm\#$37& &     &      &  \\
\hline
Soft & 0.05  & 1   & Soft & Atomic cooling \\
$\rm\#$55  &   &     &      &     \\
\hline
Massive& 0.5   & 0.1 & Hard & Massive \\
$\rm\#$186&    &     &      & ($V_{\rm c}=76.5$ km/s) \\
\hline
Radio& 0.05  & 1   & Hard & Atomic cooling \\
&    &     &      &     \\
\hline

\end{tabular}

\caption {Parameters of the models that we consider: star formation
  efficiency $f_*$, X-ray efficiency of X-ray sources $f_X$, spectral
  energy distribution (SED) of X-ray sources, and minimum circular
  velocity $V_{\rm c}$. The first four models are taken from
  \citet{Cohen:2017} [case numbers from there are indicated]; these
  all have a total CMB optical depth $\tau$ = 0.066. The Radio model
  has a radio background amplitude $A_{\rm r}=4.2$ (measured at the
  central EDGES frequency of 78~MHz, and corresponding to $0.22\%$ of
  the CMB at 1.42~GHz) and $\tau$ = 0.0737.}

\label{table:casesparam}

\end{table}

\subsection{Angular resolution, thermal noise, and smoothing}
\label{Sec:thermal}

We generated mock signals that correspond to observations with the SKA
(i.e., the low-frequency instrument of the phase-one SKA), in terms of
various resolutions and the expected thermal noise for each. It is
interesting to consider various resolutions (not only the highest
achievable SKA resolution) since low resolution images have
significantly lower noise. To create these mock 21-cm maps, we used
the following procedure [\citet{Koopmans:2015}; also L.\ Koopmans,
  personal communication].

We adopted the reasonable approximation of a Gaussian point-spread
function (PSF). Thus we used the 3~Mpc voxels (i.e., 3-D pixels) in
our simulation box but for each resolution we smoothed the signal map
with a two-dimensional Gaussian with full-width at half max (FWHM) of
$2R$, where $R$ is the smoothing radius. We illustrate our results
with three values of $R$, 10, 20, and 40~Mpc. In terms of the
telescope array, the FWHM corresponds to $\sim 0.6\lambda/D$, where
$\lambda$ is the wavelength and $D$ is the diameter within which
baselines are included. Different resolutions correspond to using
different values of $D$, so the dependence of the noise on the
resolution depends on the distribution of baselines. In the frequency
direction, the voxel size was always fixed at 3~Mpc. Now, the PSF also
indicates how the thermal noise is correlated in the image. To produce
a realistic noise map, we first generated a map of independent
Gaussian random variables in each voxel with $\sigma=1$. We then
smoothed (each slice of) the map using the same two-dimensional
Gaussian with FWHM $2R$, which gave the correct angular
correlations. The map was then rescaled so that each slice has the
expected root mean square (RMS) value of the noise for the SKA, which
depends on the redshift and the smoothing radius $R$ approximately as
[\citet{Koopmans:2015}; also L.\ Koopmans, personal communication]:
\begin{equation}
\label{Eq:noise} 
\rm \sigma_{\rm thermal}=\begin{cases} a \left( \frac{1+z}{17}
\right)^b &\mbox{if } z \leq 16\ , \\a \left( \frac{1+z}{17} \right)^c
&{\rm otherwise}\ , \end{cases}
\end{equation}
where $a$, $b$ and $c$ are the numerical coefficients for each
smoothing radius given in Table~\ref{table:noisecoef} (assuming a
1000~hr integration by the SKA). Finally, the resulting noise map was
added to the signal.

\begin{table}
\centering
\begin{tabular}{ c c c c } 

\hline
$R\, \rm[Mpc]$ & 10 & 20 & 40 \\ 
\hline
$a\, \rm[mK]$ & 15 & 4.0 & 1.8 \\ 
$b$ & 3.1 & 2.7 & 2.8 \\ 
$c$ & 4.7 & 5.1 & 4.2  \\ 
\hline

\end{tabular}

\caption {The numerical coefficients for each smoothing radius, for
  thermal noise of the SKA as given by Equation~\ref{Eq:noise}.}
\label{table:noisecoef}

\end{table}

A given resolution corresponds to 2-D Gaussian smoothing with a radius
$R$, but it is also useful to consider applying additional 3-D
smoothing as a step in the data analysis. The idea is to produce a
more isotropic image, which is more conducive for measuring statistics
that are designed to probe spherically-averaged structure. Now, while
any smoothing removes some information in the map, it also smooths out
and thus lowers the thermal noise. In our results below, we have found
that the differences are small between using the images with or
without 3-D smoothing, but in most cases the 3-D smoothing increases
the signal-to-noise ratio, i.e., the noise is smoothed-out more than
the signal. This makes sense since the typical coherence/correlation
scale of the noise is $R$ (due to the PSF), while the typical scales
of the 21-cm features (due to reionization, heating, or Ly$\alpha$
coupling) are usually significantly larger. Thus as our default
procedure we did include 3-D smoothing, using a spherical top-hat with
the same smoothing radius $R$ as in the corresponding 2-D Gaussian.

\section{Statistical Methods and Results}
\label{Sec:Results}

We first show the sky-averaged (global) signal for all five models
from Table~\ref{table:casesparam} as predicted from the
simulation. All five curves show the same general behavior of a deep
absorption dip, as is the case for all reasonable models
\citep{Cohen:2017}. Three models are especially similar in their
timing: the Standard, Soft, and Radio models have relatively early
Ly$\alpha$ coupling and X-ray heating, resulting in a peak absorption
at $z \sim 18-19$, followed by a rise to emission ($\Tb > 0$) before
the drop to zero due to reionization. On the other hand, the
Low-Efficiency and Massive models have much later star formation, so
that Ly$\alpha$ coupling is delayed and X-ray heating overlaps with
reionization and does not manage to lead to emission.  Comparing the
Soft model to the Standard one, the heating phase starts earlier in
the Soft model, leading to an earlier rise from the absorption
trough. The Radio model has a very deep $\rm{Ly}\alpha$ minimum due to
the excess radio background.

\begin{figure}
  \centering
          \includegraphics[width=3.2in]{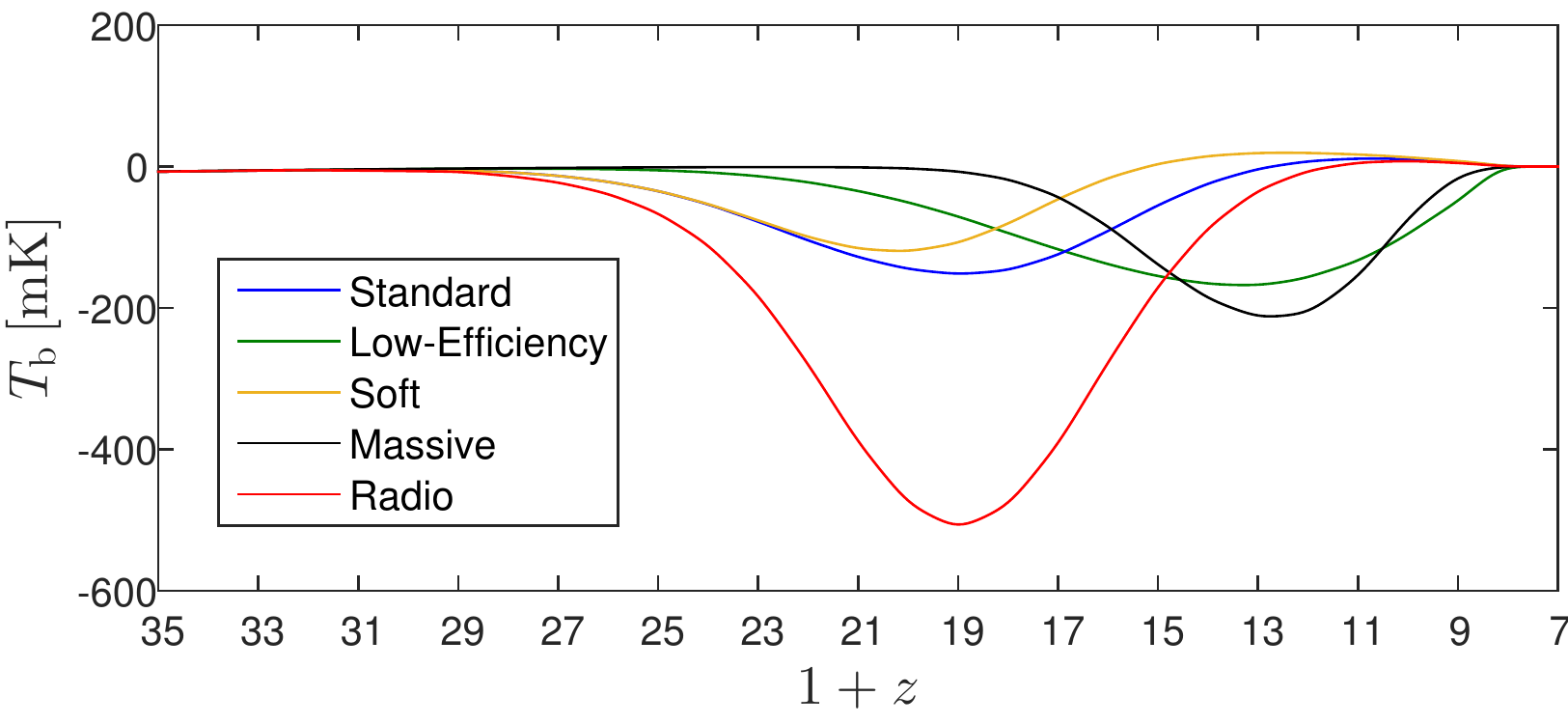}
\caption{The global 21-cm signal as a function of redshift for our
  five models, Standard (blue), Low-Efficiency (green), Soft (orange),
  Massive (black), and Radio (red).}
\label{fig:global}
\end{figure} 

\subsection{Histograms and quantiles}

\label{s:thresh}

\begin{figure*}
\centering
\includegraphics[width=3.1in]{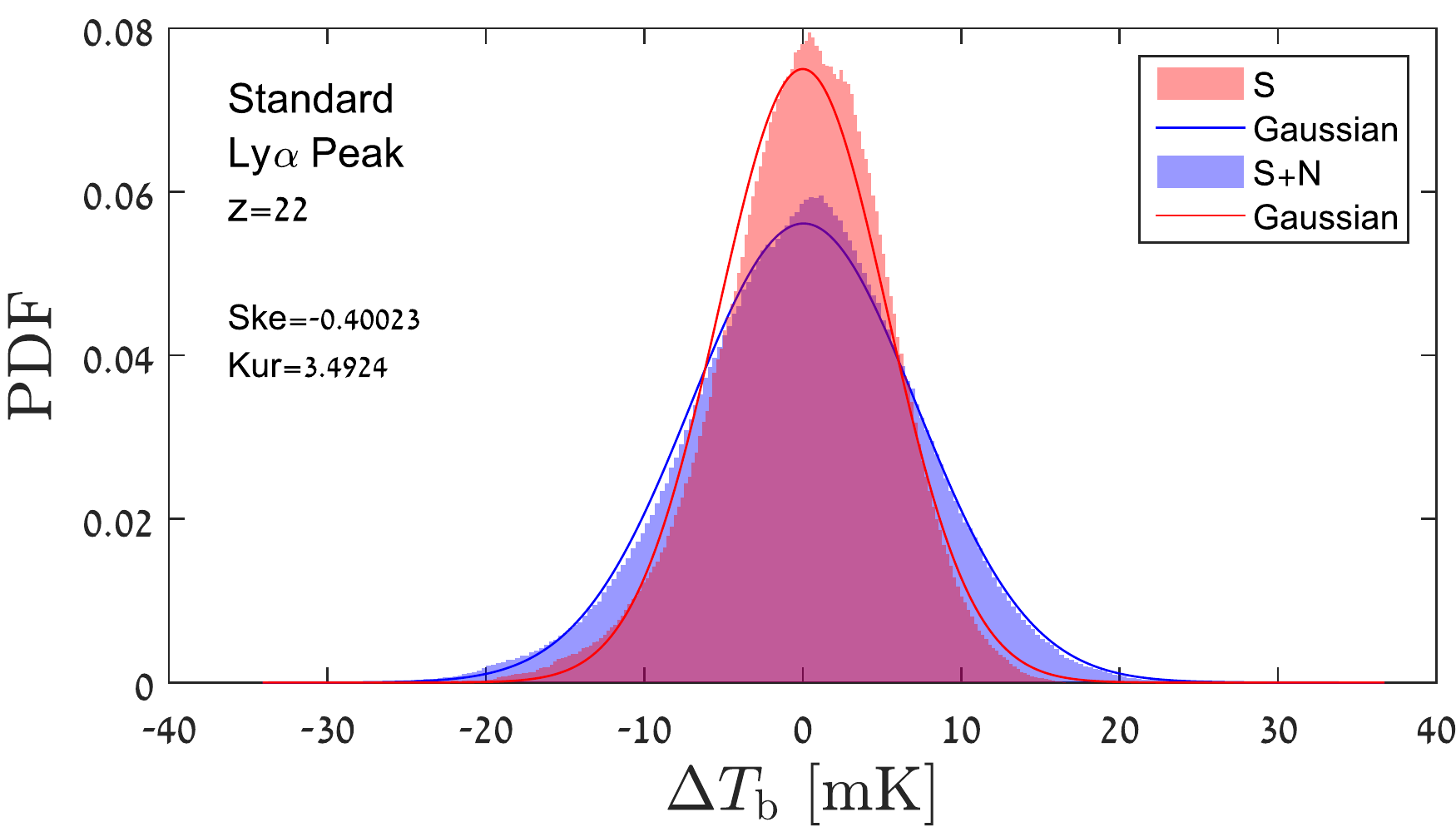}\hspace{0.2in}
\includegraphics[width=3.1in]{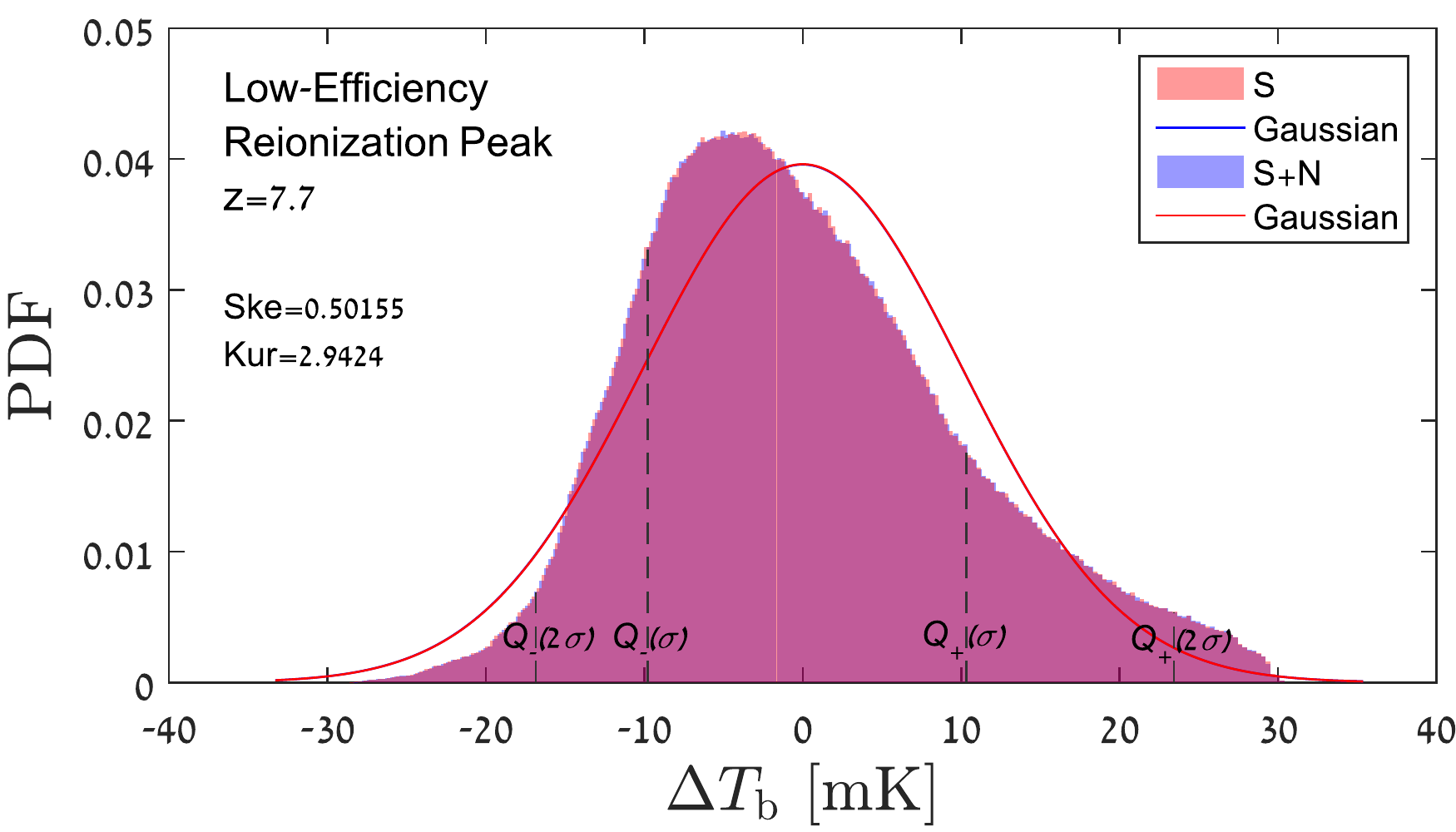}

\caption{Probability distribution functions (PDFs) of the noiseless
  signal (peach histogram; red curve) and the signal with added noise
  (light blue histogram; blue curve), assuming an $R=20$~Mpc angular
  resolution (plus 3-D smoothing). Each solid curve corresponds to a
  Gaussian PDF with the same mean and variance as the histogram with
  matching color, for comparison. Left panel: Standard model at the
  Ly$\alpha$ peak (for definitions of the various peaks, see
  section~\ref{s:qave} below). Right panel: Low-Efficiency model at
  the Reionization peak. The skewness and kurtosis values of the
  signal-only histograms are also noted; these measures are discussed
  in detail in the last two subsections of
  section~\ref{Sec:Results}. The added SKA thermal noise is discussed
  in section~\ref{Sec:thermal}. In the right panel, the quantiles
  $Q_+$ and $Q_-$ are shown for the signal, for $t=1\sigma$ as well as
  $t=2\sigma$. Note that in the right panel the noise is negligible
  due to the lower redshift, making the two histograms in it nearly
  identical.}
\label{fig:Distributions}
\end{figure*}

Our statistical tools are mostly based on histograms of the 21-cm
signal map, i.e., the probability distribution function $p(T_b)$ of
the 21-cm intensity (brightness temperature $T_b$) in voxels,
normalized to a total area of unity. As the variable we use $\Delta
T_b$, which is $T_b$ measured relative to the mean temperature at the
same redshift, since interferometers do not measure the zero point.
Figure~\ref{fig:Distributions} shows two examples of such histograms
for separate models and cosmic times. The distributions are clearly
non-Gaussian, and one of the main features we focus on is the obvious
asymmetry. The shape of the asymmetry depends in a complex way on the
astrophysical processes and parameters; these examples illustrate
opposite signs of the skewness.

In what follows, we use the cumulative distribution function (CDF) of
the signal, either the upper portion:
\begin{equation}
F_+(\Delta T_b) \equiv \int_{\Delta T_b}^\infty p(\Delta \hat{T}_b)\, d\Delta \hat{T}_b\ ,
\end{equation}
or the lower portion:
\begin{equation}
F_-(\Delta T_b) \equiv \int_{-\infty}^{\Delta T_b} p(\Delta \hat{T}_b)\, d\Delta \hat{T}_b\ .
\end{equation}
Note that $F_+(-\infty)=F_-(\infty)=1$. We measure characteristic
brightness temperatures as thresholds at certain values of the
CDF. This is the inverse function of the CDF (also called the quantile
function $Q$, which here has units of mK). For a given fraction $f$ of
the total probability, we have an upper threshold $Q_+(f)$ so that a
fraction $f$ of the probability lies at temperatures above $Q_+(f)$,
and similarly a lower threshold $Q_-(f)$. They are defined so that
\begin{equation}
F_+(Q_+(f)) = f\ ,
\end{equation}
and
\begin{equation}
F_-(Q_-(f)) = f\ .
\end{equation}
For the probability fractions we use characteristic thresholds $t$
based on the cumulative probability of a normal distribution, measured
in units of the standard deviation $\sigma$. For instance, we define
$Q(t=1\sigma) \equiv Q(f=15.9\%)$, where this holds for both $Q_+$ and
$Q_-$. Note that $Q_+$ and $Q_-$ are defined to be one-sided so we use
the corresponding one-sided fractions of a Gaussian (e.g., $f=15.9\%$
for $t=1\sigma$, not $f=31.7\%$). More generally, the relation between
$f$ and $t$ is given by
\begin{equation}
\label{Eq:Threshold} 
f(t)=\frac{1}{2} {\rm erfc} \left(\frac{\textit{t}}{\sqrt{2}}\right)\ ,
\end{equation}
where $t$ is measured in units of
$\sigma$. Table~\ref{table:Thresholds} lists the values of various
thresholds that we use below along with their corresponding
percentiles, according to eq.~\ref{Eq:Threshold}. Note that for a Gaussian distribution,
$Q_+(t) = -Q_-(t) = t \sigma$.

Quantiles for one case are shown in the right panel of
Figure~\ref{fig:Distributions}. In this case, $Q_+$ and $Q_-$ have
nearly the same magnitude at $1\sigma$; while $Q_-$ is closer than
$Q_+$ to the peak of the PDF as well as to its median, we have defined
$Q_+$ and $Q_-$ as they are measured in 21-cm images, i.e., relative
to the cosmic mean brightness temperature. At the $2\sigma$ threshold
the difference becomes clear, with the higher $|Q_+|$ reflecting the
broader tail at high brightness temperature. In the Low-Efficiency
model shown here during reionization, the intergalactic medium is
still cold, so that the high $T_b$ tail corresponds to regions that
are mostly reionized (though not completely so, due to the smoothing
of the map, which mixes ionized bubbles with nearby pixels that are
still partly neutral).

\begin{table}
\centering
\begin{tabular}{ r c l r } 

\hline
$t$ & $f(t)$ & $1-f(t)$ & $\rm{N_{vx}}$\\ 
\hline
0.5$\sigma$  & 30.9\%   & 69.1\%    & 647,000 \\ 
1$\sigma$   & 15.9\% & 84.1\%    & 333,000 \\ 
1.5$\sigma$ & 6.7\%  & 93.3\%  & 140,000 \\ 
2$\sigma$    & 2.28\%  & 97.72\%  & 47,700  \\ 
2.5$\sigma$  & 0.62\% & 99.38\% & 13,000  \\ 
3$\sigma$   & 0.135\% & 99.865\% & 2,830   \\
\hline

\end{tabular}

\caption {List of thresholds used in this paper along with the
  corresponding percentiles of the normal distribution. $\rm{N_{vx}}$
  denotes the actual number of voxels corresponding to the fraction
  $f(t)$, for a 128$^3$ voxel simulation box as used here.}
\label{table:Thresholds}

\end{table}

\subsection{Radial profiles}

In most of our analysis below, we focus on the PDF of $T_b$ values and
various derived statistics as laid out in the previous
subsection. This approach brings out non-Gaussianity most clearly, and
makes thermal noise especially easy to deal with. However, there is
additional spatial information that can be derived from the 21-cm
map. We briefly give an example of this here.

\begin{figure*}
\centering
\includegraphics[width=7in]{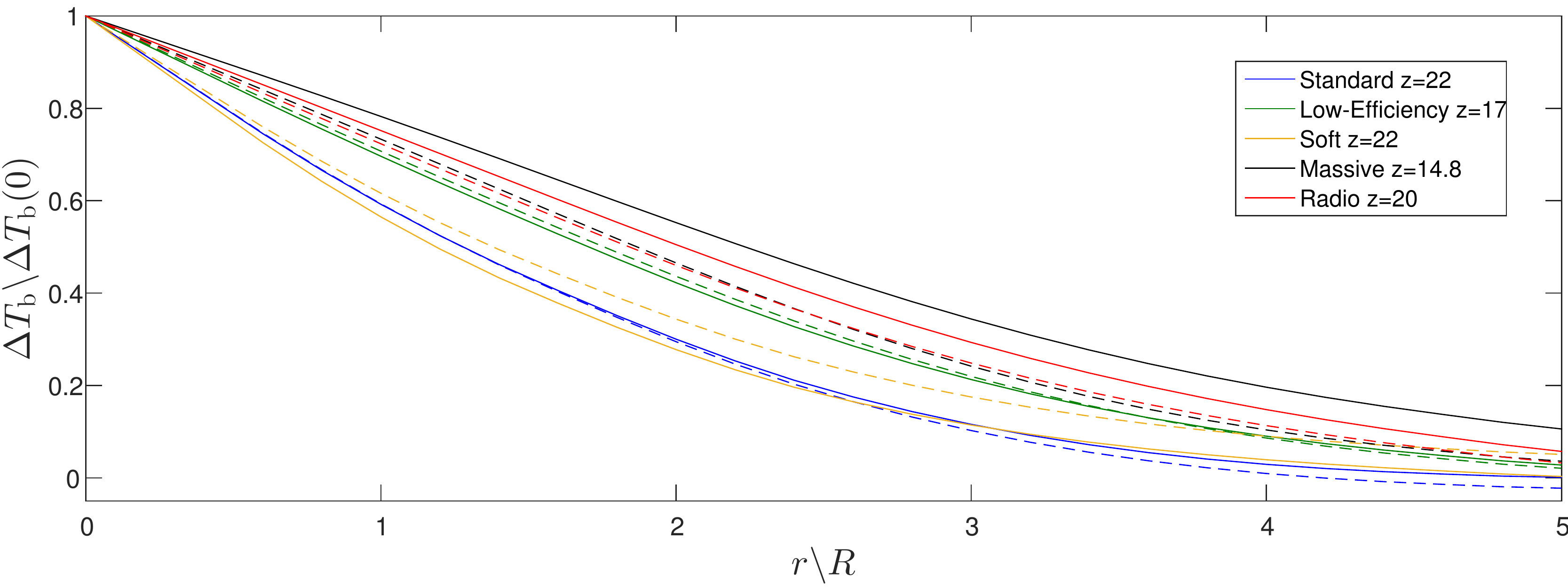}

\caption{Normalized radial $\Tb$ profiles for the various models
  (Table~\ref{table:casesparam}), with $R=20$~Mpc and at redshifts
  corresponding to the $\rm{Ly}\alpha$ peak (for definitions of the
  various peaks, see section~\ref{s:qave} below). Standard - blue,
  Low-Efficiency - green, Soft - orange, Massive - black, Radio -
  red. Solid lines show the average profile around the 15.9\% highest
  voxels in the map, and dashed lines show the average profile around
  the 15.9\% lowest voxels.}

\label{fig:profiles}
\end{figure*}

We can use the thresholds to explore what roughly corresponds to
radial profiles around temperature peaks. Specifically, we found the
average profiles around the voxels with the highest or lowest values
of $\Delta \Tb$. From this we can examine the contribution of various
spatial scales to the fluctuations and also look for asymmetry (and
thus non-Gaussianity) by comparing the highest and lowest
voxels. Since we wanted average spherical profiles, in order to select
the voxels we used as before the 3-D spherically averaged $\Tb$ around
each voxel. As an example, we chose $R=20$~Mpc and used the 15.9\%
highest and lowest voxels (corresponding to $t=1\sigma$ in the
previous subsection). To find the profiles, at each distance $r$ we
found the volume-averaged smoothed signal in the shell that includes
points at distances between $r-R$ and $r+R$ from the central voxel. For
$r=0$ we simply used the spherical average out to radius $R$. Finally,
the profiles of each group (highest or lowest pixels) were stacked to
produce an average profile for each group. Figure~\ref{fig:profiles}
illustrates the resulting profiles (shown normalized, relative to
$r=0$) for all five models at the $\rm{Ly}\alpha$ peak. Differences
between the profiles of the highest and lowest pixels are visible for
all models, i.e., there is clear asymmetry. Also, different models show
different characteristic scales for the drop of the profile. For
example, the profile that declines most slowly (i.e., shows the
strongest large-scale correlations) corresponds to the Massive model,
where the halos are massive, rare, and more highly biased than in the
other models.

\subsection{Quantiles and noisy maps}

From here on, we consider the quantiles at various thresholds as
defined in section~\ref{s:thresh}. At each threshold level $t$, we
find $Q_+(t)$ and $Q_-(t)$, which measure the brightness temperature
above or below the mean which describes a fraction of the map
corresponding to that threshold. These quantities probe the magnitude
of the positive and negative fluctuations, and the choice of $t$ gives
us controls: a higher threshold $t$ corresponds to probing rarer
fluctuations, while a lower threshold is more robust and less
sensitive to noise, especially to outliers in the data. Standard
statistical measures average over the entire distribution and do not
offer such flexibility. As we show, we can reconstruct the standard
non-Gaussian statistics with quantiles, plus look for additional
measures.

For a Gaussian distribution, a quantile at a given threshold value
would give a brightness temperature that is a fixed multiple of the
standard deviation $\sigma$ of the distribution. Thus, in general,
what a quantile measures is roughly (a multiple of) the standard
deviation. Now, in general, the total variance of the noisy signal
equals the sum of the signal variance and noise variance (assuming
that they are independent). This leads us to use a simple procedure
for correcting the measured quantiles from our mock data for the
effect of noise. The estimated signal is taken as
\begin{equation}
  \label{e:est}
S_{\rm{est}}=\sqrt{({S+N_1})^2-{N_2}^2}\ ,
\end{equation}
where $S_{\rm{est}}$ refers to the estimated signal (either $Q_+$ or
$Q_-$ at some threshold $t$), $S+N_1$ is the measured signal from a
21-cm image with signal plus thermal noise, and $N_2$ is the same
quantity measured from a noise-only 21-cm image, using noise $N_2$
generated independently from $N_1$. Thus, we assume that in the data
analysis the statistical properties of the thermal noise are known
(but not the particular instance that is included in the measured
data). We note that it is not obvious that this noise-correction
procedure, which is based on variances, applies exactly to quantiles
even for non-Gaussian signals. In practice, though, we find that it
works very well, and we thus conclude that this simple
noise-correction property is an important advantange of working with
quantiles.

The estimation in all plots was made up to redshift 27 which
approximately corresponds to the SKA's lowest measured frequency of
50~MHz. Note that the signal maps were generated with a redshift
resolution of $\Delta z=0.1$ up to redshift 15 and $\Delta z=1$ above
this, for all models except for the Radio model where we used
resolution of $\Delta z=1$ for all redshifts.

\subsection{Quantile average compared to variance}

\label{s:qave}

\begin{figure*}
\centering
\includegraphics[width=7in]{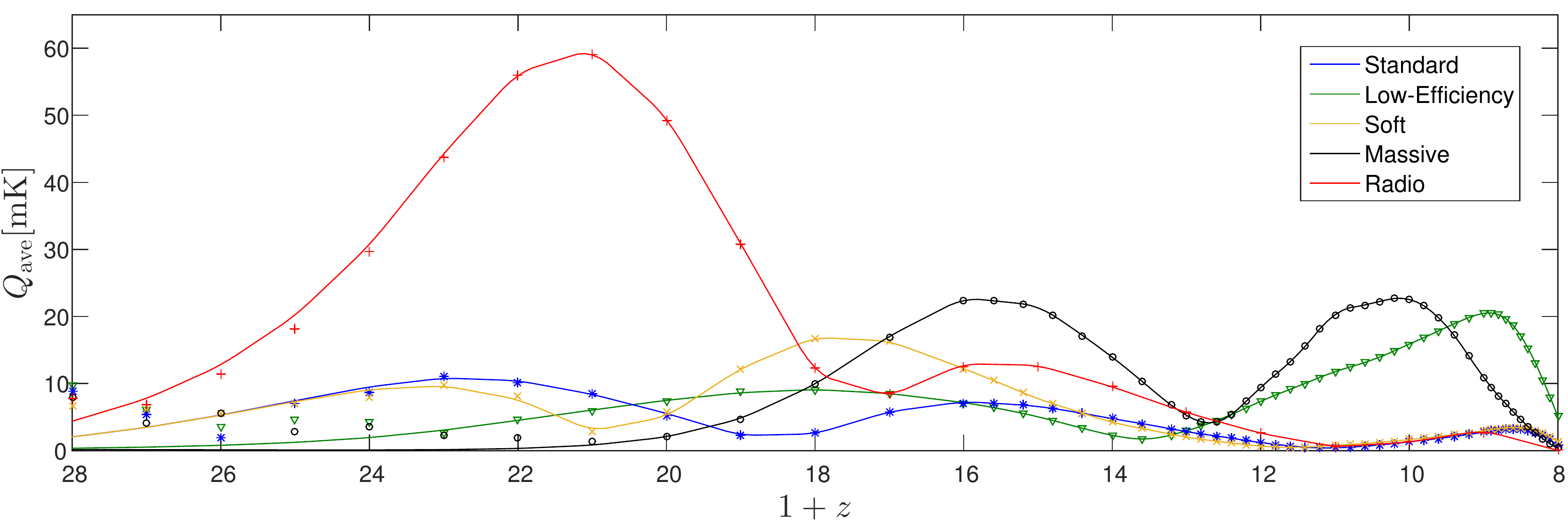}
\includegraphics[width=7in]{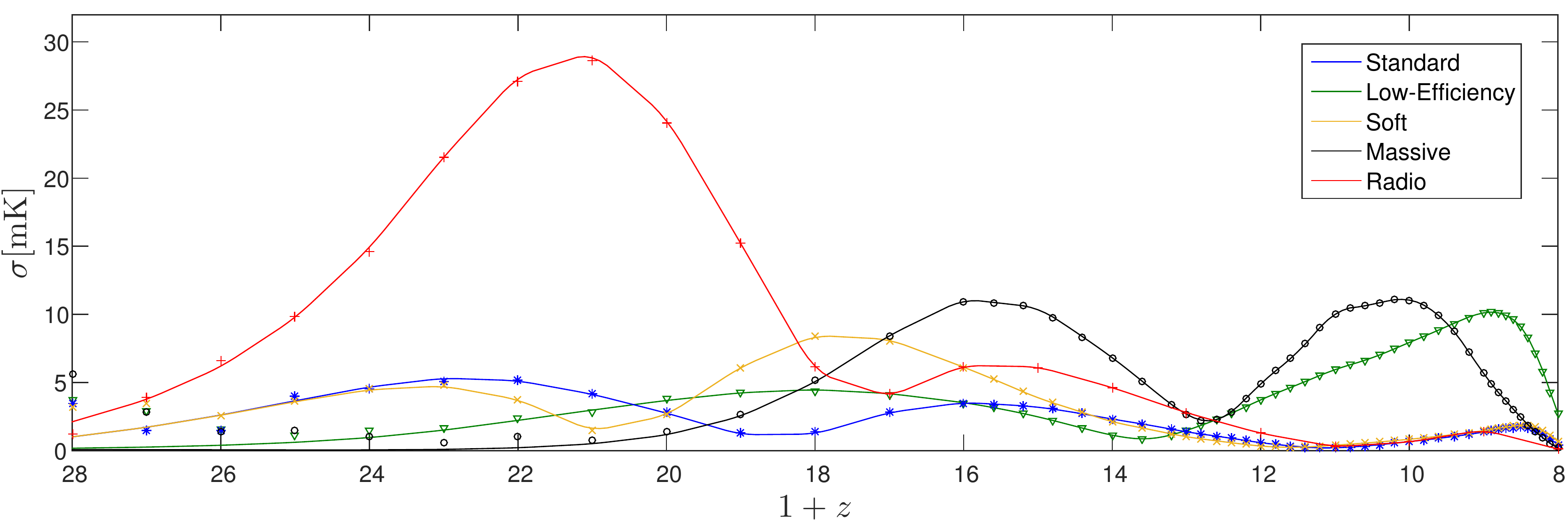}

\caption{Statistics measured versus redshift, from a noise-less 21-cm
  image (curves) compared to the noise-corrected estimated statistics
  from a 21-cm image with added noise (points), with $R=20\, \rm Mpc$
  (2-D plus 3-D smoothing). Top panel: quantile average of high- and
  low-end 2$\sigma$ thresholds. Bottom panel: regular standard
  deviation of the PDF. All of our five models are shown. Note that
  the density of points changes according to the output redshift
  resolution of each simulated model.}

\label{fig:AveAll}
\end{figure*}

The first quantile measure we looked at is the average (in absolute
value) of the high- and low-end quantiles, i.e.,
\begin{equation}
\label{e:ave}
  Q_{\rm ave}(t) \equiv \frac{|Q_+(t)| + |Q_-(t)|}{2}\ .
\end{equation}
Note that, by their definitions, $Q_+$ is positive and $Q_-$ is
negative (not necessarily for all possible distributions, but this is
the case for all realistic ones). This quantity would equal $t$ times
$\sigma$ for a Gaussian distribution, and more generally it
corresponds to estimating the distribution's standard deviation
(except for the factor of $t$). By averaging the two ends we ignore
any asymmetry and get an accurate estimate of the symmetric part. As
our main configuration we use a 2$\sigma$ threshold and $R=20\, \rm
Mpc$. We could get a similar result here with the more natural
1$\sigma$, but we prefer to keep the same choice later when we look at
the difference, and that signal happens to nearly vanish for a
1$\sigma$ threshold (see Figure~\ref{fig:DifCases},
below). Figure~\ref{fig:AveAll} shows the average for all five models
as a function of redshift with the above main configuration
parameters, with the regular standard deviation of the PDF shown for
comparison. As with the quantiles, the variance estimation from the
noisy map was corrected for noise by subtracting the variance of an
independent noise map:
\begin{equation}
\sigma_{\rm{est}}=\sqrt{\rm{Var}(\textit{S+N}_1)-\rm{Var}(\textit{N}_2)}\ .
\end{equation}

From the plot, the quantile average accurately measures the standard
deviation (times a factor of 2 in this case, i.e., our main
configuration). Compared to the noise-less image, the noise-corrected
estimation from the noisy map performs very well, nearly up to the
highest redshifts considered, for both the quantile average and the
standard deviation statistic, and for all models. The exceptions are
redshifts at which the signal drops near zero for some models.

\begin{figure*}
\centering
\includegraphics[width=3.1in]{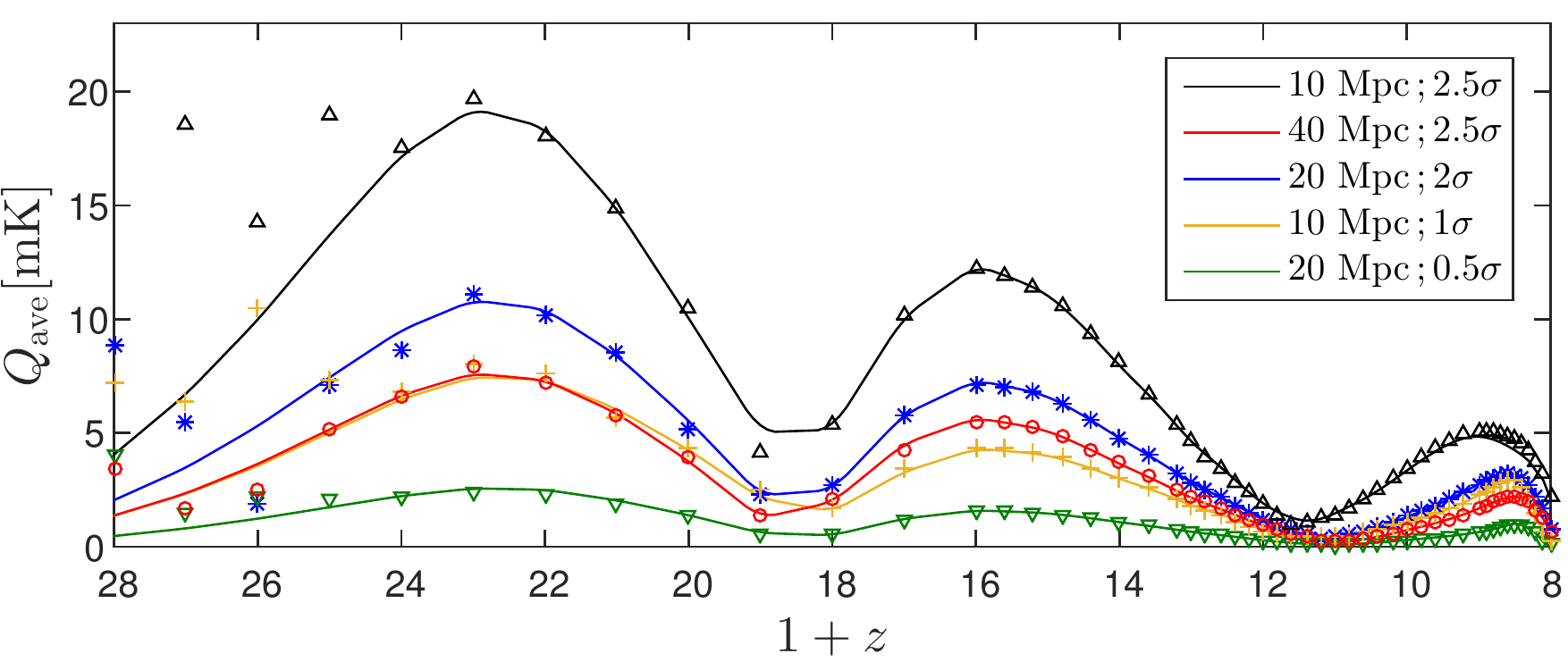}\hspace{0.2in}
\includegraphics[width=3.1in]{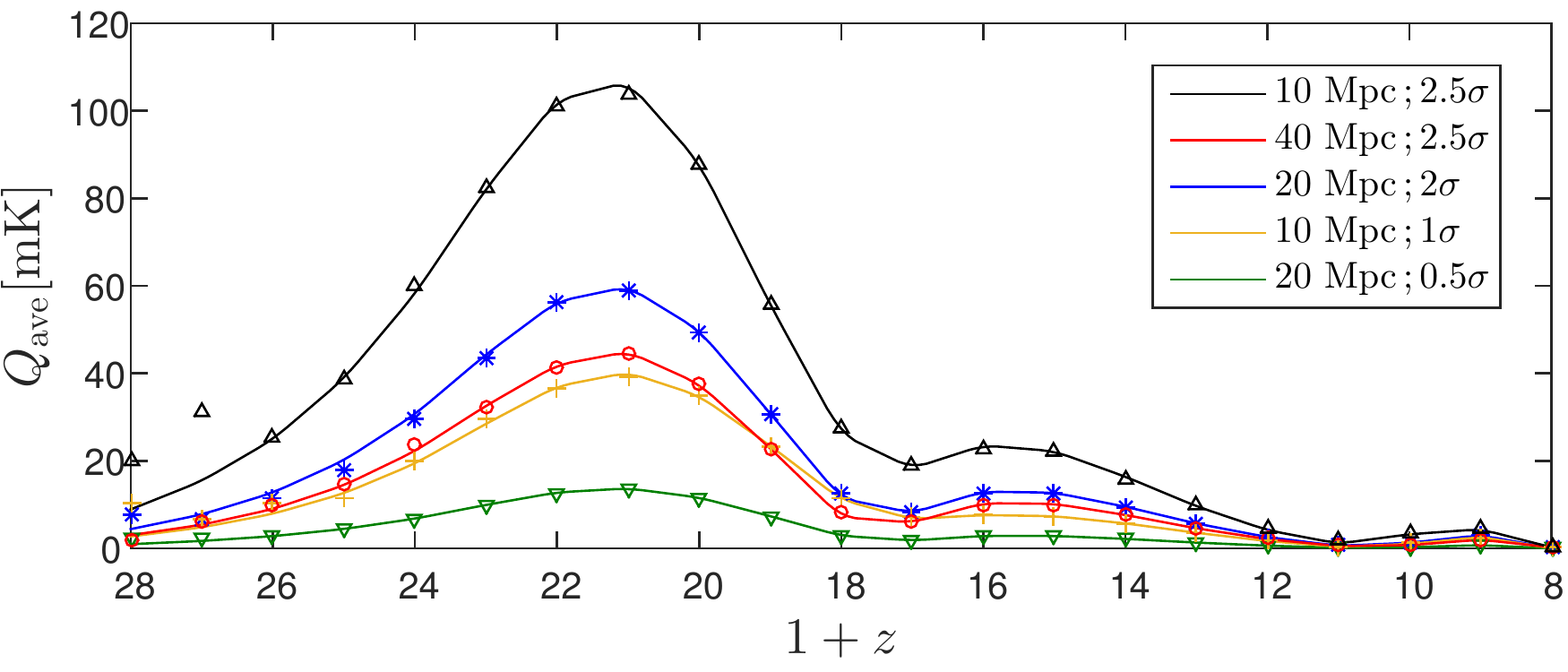}

\caption{Quantile average of high- and low-end thresholds of the
  brightness temperature distribution as a function of redshift, for
  several configurations of thresholds and smoothing radii $R$. Left
  panel: Standard model, Right panel: Radio model. As in the previous
  figure, curves are from the noise-less signal and symbols show the
  estimated statistic from the map with added noise.}

\label{fig:AveCases}
\end{figure*}

As noted above, our ability to control the threshold and smoothing
radius allows us to look at different parts of the temperature
distribution and at various scales (similar to what we do when using
the power spectrum), and to manipulate the magnitudes of the signal
and noise since smoothing affects them differently. Figure
\ref{fig:AveCases} illustrates the effect of using different parameter
configurations for the Standard and Radio models. For high threshold
and low $R$ we get the biggest magnitude, but as can be seen for the
Standard model (left panel), with this choice the estimation fails for
$z > 22$ and also becomes inaccurate below 8. The Radio model has a
particularly strong signal and thus yields more accurate estimates at
the highest redshifts. We conclude that having the option to control
the two parameters that are varied here has the potential to yield
more information from the analysis of a real dataset.

Most of our plots in this paper are presented as functions of
redshift. However, as noted above, when we wish to select particular
milestone redshifts, we define them phenomenologically using the
(mock) estimated signal. Specifically, we use the redshifts where our
main measure of the signal, the quantile average, achieves a peak
value (i.e., a local maximum). As seen from Figure~\ref{fig:AveAll},
from high to low redshift, in each model we have a Ly$\alpha$ peak, a
Heating peak, and a Reionization peak (except that there is no Heating
peak in the Massive and Low-Efficiency models).

\subsection{Threshold dependence}

\label{sec:thresh}

\begin{figure*}
\centering
\includegraphics[width=3.1in]{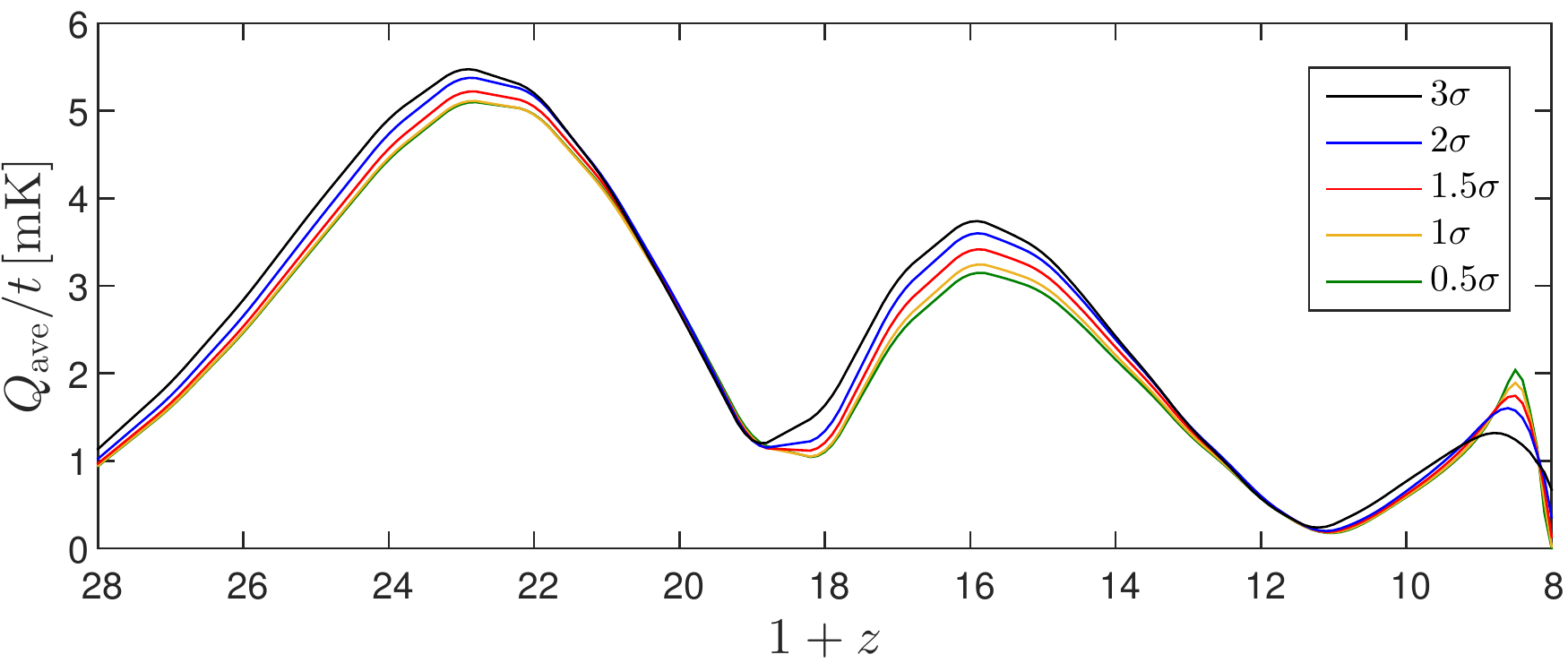}\hspace{0.2in}
\includegraphics[width=3.1in]{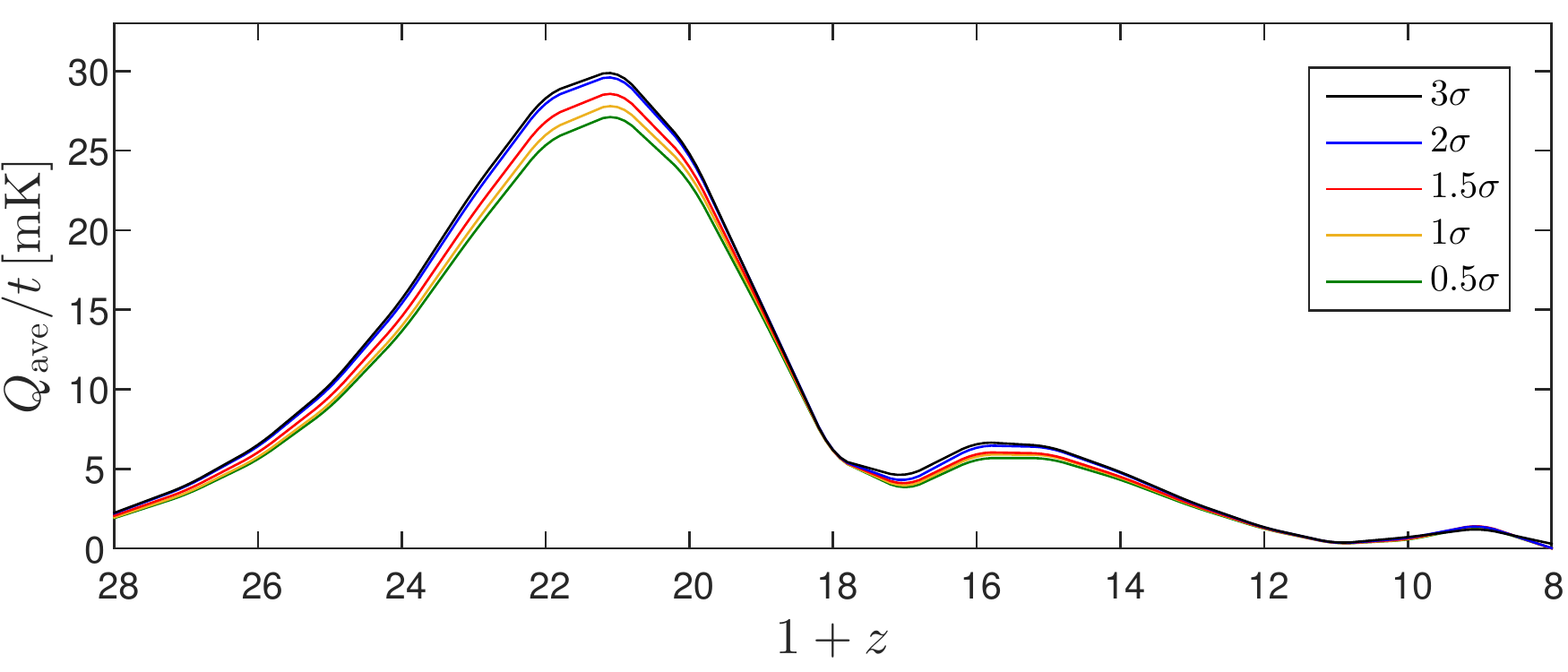}

\includegraphics[width=3.1in]{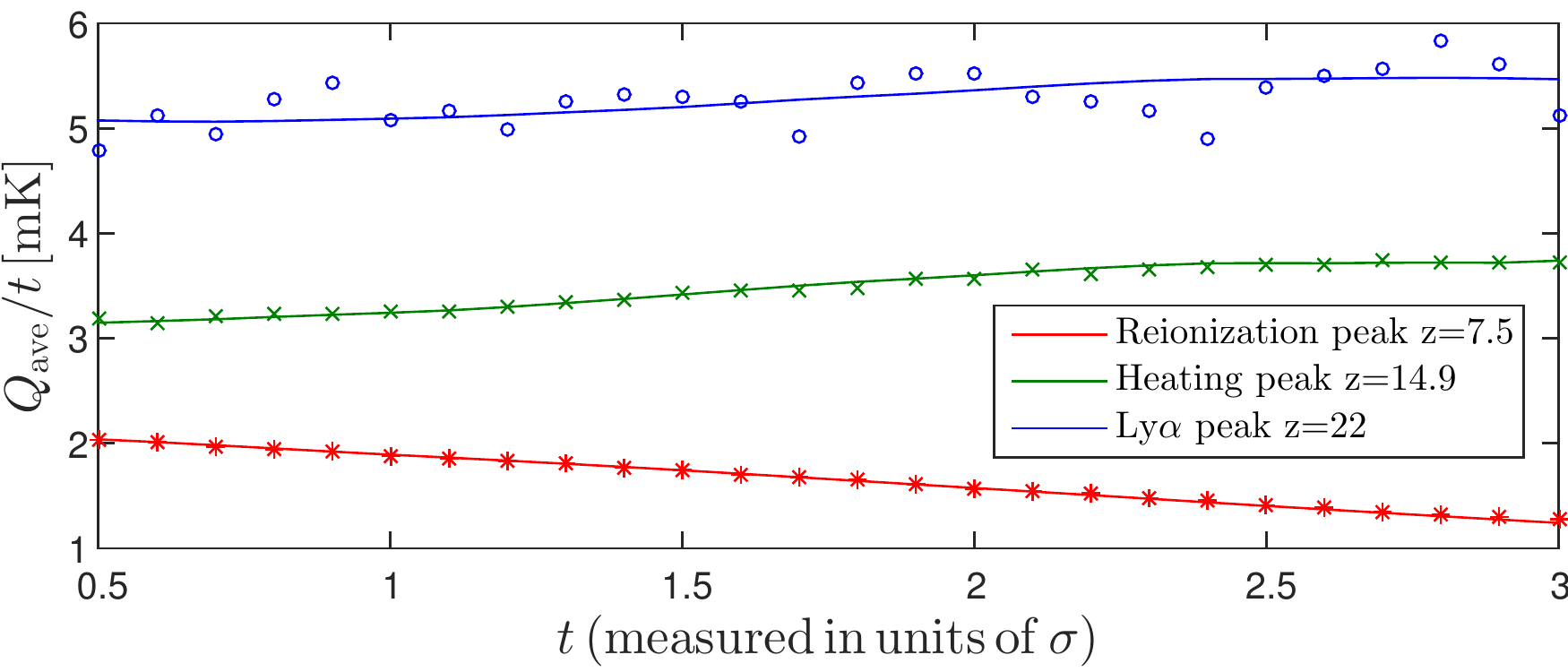}\hspace{0.2in}
\includegraphics[width=3.1in]{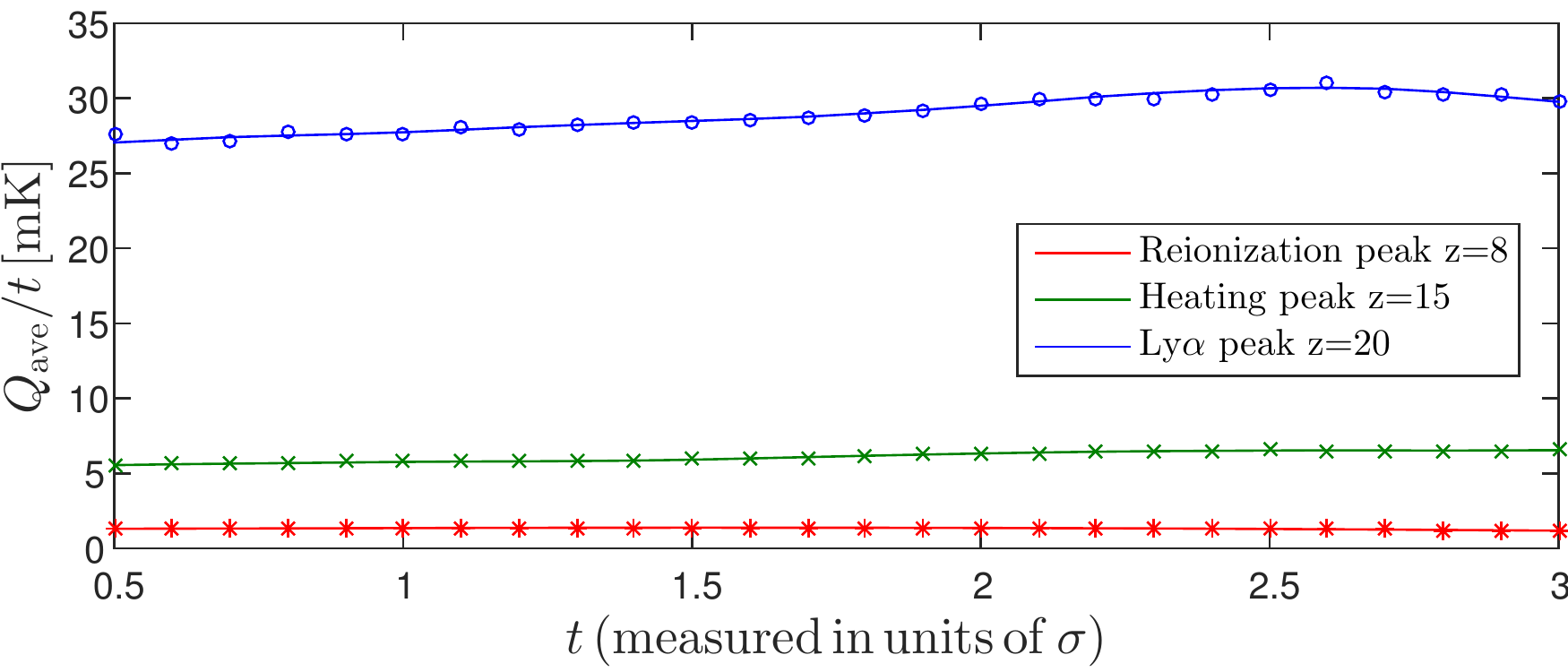}

\includegraphics[width=3.1in]{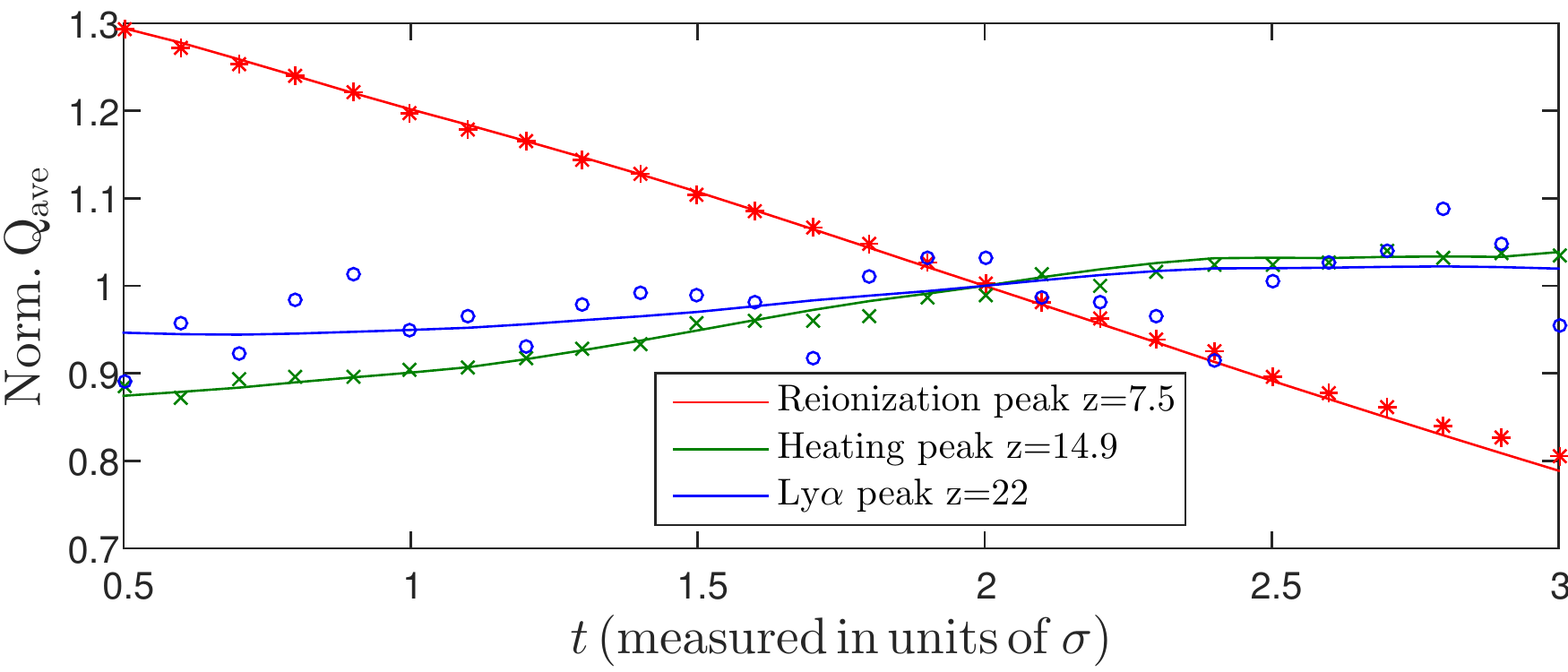}\hspace{0.2in}
\includegraphics[width=3.1in]{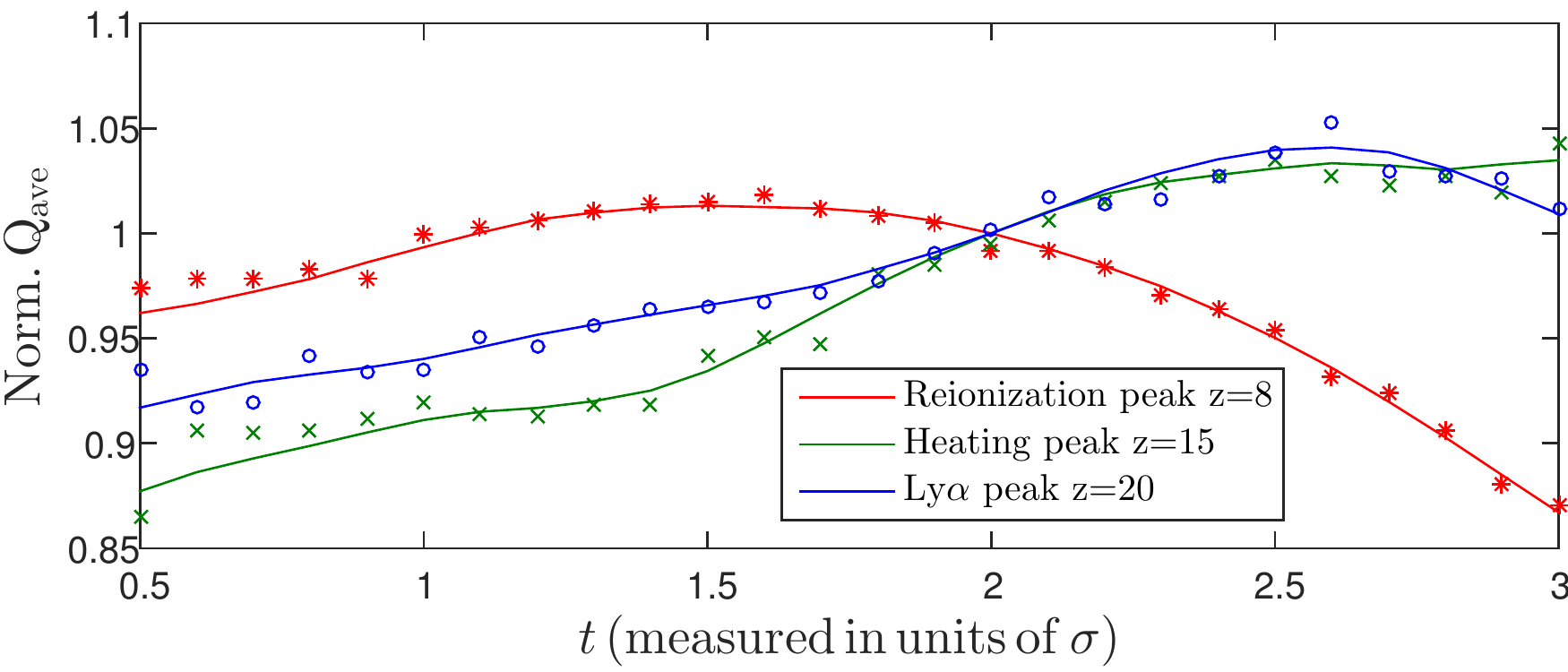}

\caption{Threshold dependence of the average signal. Left panels:
  Standard model plots, right panels: Radio model plots. Top panels:
  normalized quantile average (i.e., divided by the threshold $t$) as
  a function of redshift. Middle panels: normalized quantile average
  as a function of threshold for 3 milestone peak redshifts. Bottom
  panels: normalized (twice) curves - each curve from the middle panel
  was divided again by its value at threshold $t=2$. Symbols in the
  various panels show the same plotted statistic as estimated from
  data with added thermal noise.}

\label{fig:Threshold}
\end{figure*}

The quantiles that we have defined can be used to directly compare the
measured PDF to a Gaussian distribution, by varying the threshold and
normalizing to a Gaussian. As the first step, we calculated the
quantile-average curves (defined as in Figure~\ref{fig:AveCases} but
for a fixed $R=20\, \rm Mpc$) and normalized them according to the
threshold (e.g., the 2$\sigma$ curve was divided by 2). The resulting
curves, shown in the top panel of Figure~\ref{fig:Threshold}, would
lie exactly on top of each other for a pure Gaussian distribution.
For the simulated (noise-less) 21-cm signal there are differences, an
indication of non-Gaussianity. Note that the estimated signal from
noisy maps is not plotted here since the points would be very crowded;
there errors were illustrated in the previous two figures, and the
normalization by a constant does not change the relative errors of the
estimation.

The differences between the normalized curves are largest mostly near
the cosmological milestone redshifts. We focus on these special
redshifts in the other two panels. The middle panel shows the
normalized quantile average at each redshift, as a function of the
threshold level $t$. We bring out the variation more clearly in the
bottom panel, where we have applied yet another normalization
according to the value of each curve at the 2$\sigma$ threshold. In
these two panels, a Gaussian distribution would give a flat horizontal
line. The non-Gaussian signature is strongest during reionization, but
all the curves exhibit interesting behavior. The symbols, which
represent the same estimated statistics from the noisy signal, show
that the SKA thermal noise usually does not prevent this
non-Gaussianity from being measured; at the Ly$\alpha$ peak, the
measurement is rather noisy in the Standard model, but the stronger
signal in the Radio model allows an accurate measurement also at
$z=20$. Another interesting feature is that in the Radio model the
curves are not monotonic as they are in the Standard model. We relate
these measures of non-Gaussianity from the symmetric quantile-average
to the kurtosis in section~\ref{sec:kur}; but first we move on to the
asymmetry of the positive and negative brightness temperature
fluctuations.

\subsection{Quantile difference and skewness}

\begin{figure*}
\centering
\includegraphics[width=7in]{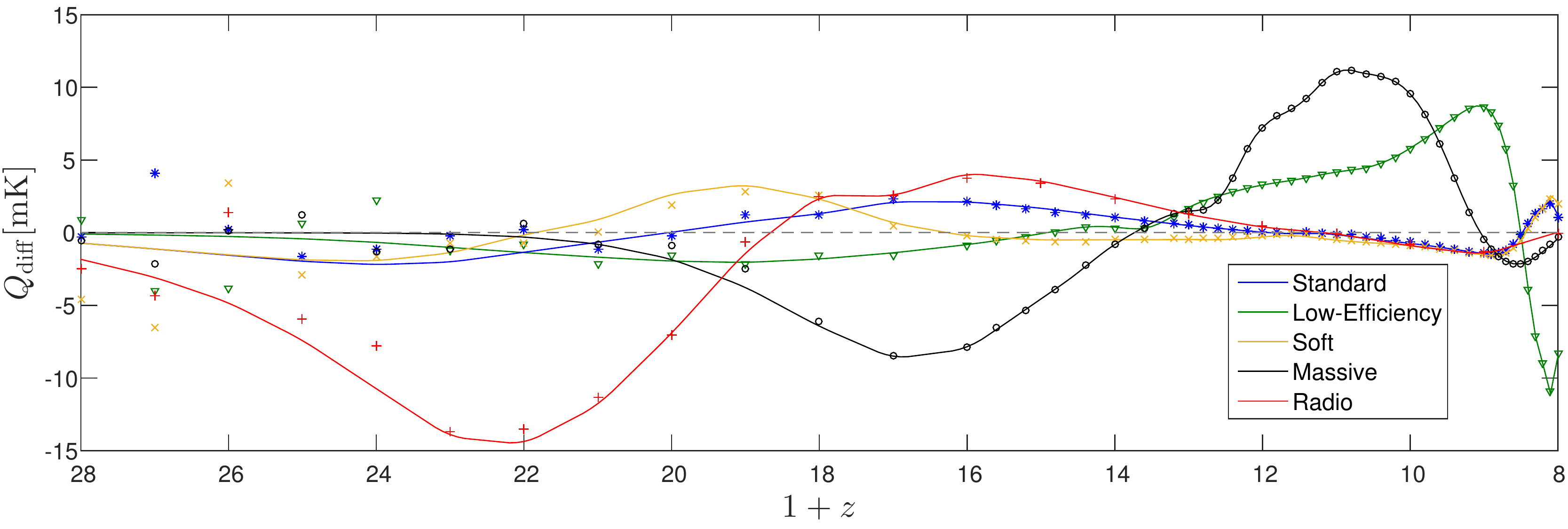}
\includegraphics[width=7in]{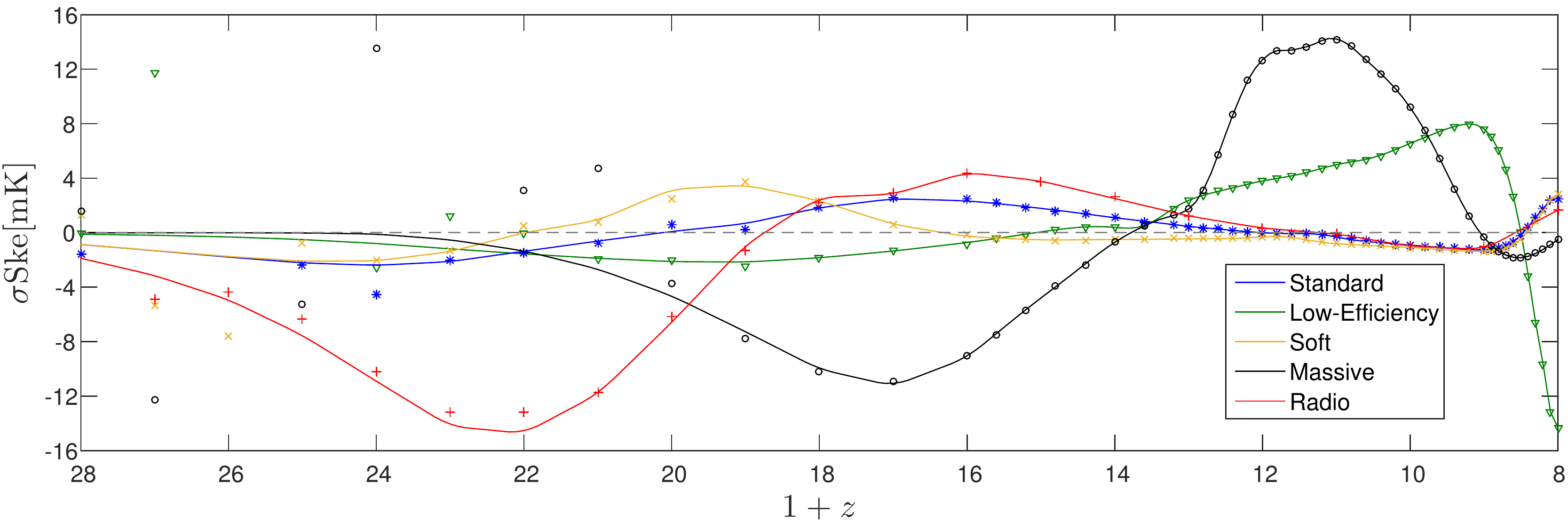}

\caption{Top panel: Quantile difference of the 21-cm fluctuations
  using 2$\sigma$ thresholds with $R=20\, \rm Mpc$, shown as a
  function of redshift for all 5 models. Bottom panel: Skewness times
  $\sigma$, as a function of redshift, for the same five models. The
  symbols represent the estimated statistic in each case from the map
  with added noise.}

\label{fig:DifAll}
\end{figure*}

We now probe the asymmetry of the PDF using the difference between the
high- and low-end quantiles, i.e.,
\begin{equation}
\label{e:diff}
Q_{\rm diff}(t) \equiv |Q_+(t)| - |Q_-(t)|\ .
\end{equation}
This quantity can be compared with the standard measure of
non-Gaussian asymmetry, namely the distribution's skewness given by
\begin{equation}
  \rm{Ske}(\textit{X})=E[(\textit{X}-\mu)^3]/\sigma^3\ ,
\end{equation}
where $\mu$ refers to the mean value of $X$ which in our case equals
zero. Both of these measures of asymmetry would equal zero for a
Gaussian PDF, and cannot be probed using the 21-cm power spectrum
(which measures the contribution of $k$-modes to the
variance). Figure~\ref{fig:DifAll} shows our quantile difference
statistic, as well as the skewness (multiplied by the measured
$\sigma(z)$ to make it have dimensions of brightness temperature), for
all five astrophysical models, with the main configuration parameters
(2$\sigma$ threshold with $R=20\, \rm Mpc$). We see that the two
statistics are quite similar (though not identical), and can be
estimated accurately from a noisy map except when the signal is low at
$z>20$. The skewness estimation from the noisy map was done using the
formula:
\begin{equation}
  \rm{Ske}_{[\rm{est}]}=\frac{\rm{Ske}(\textit{S+N}_1)\rm{Var}^{3/2}
  (\textit{S+N}_1)}{(\rm{Var}(\textit{S+N}_1)-\rm{Var}
  (\textit{N}_2))^{3/2}}\ .
\end{equation}
This is easily derived from the fact that
the Gaussian noise has zero skewness, and the skewness of the signal
is defined with respect to the variance of the (noise-less) signal.

\begin{figure*}
\centering
\includegraphics[width=3.1in]{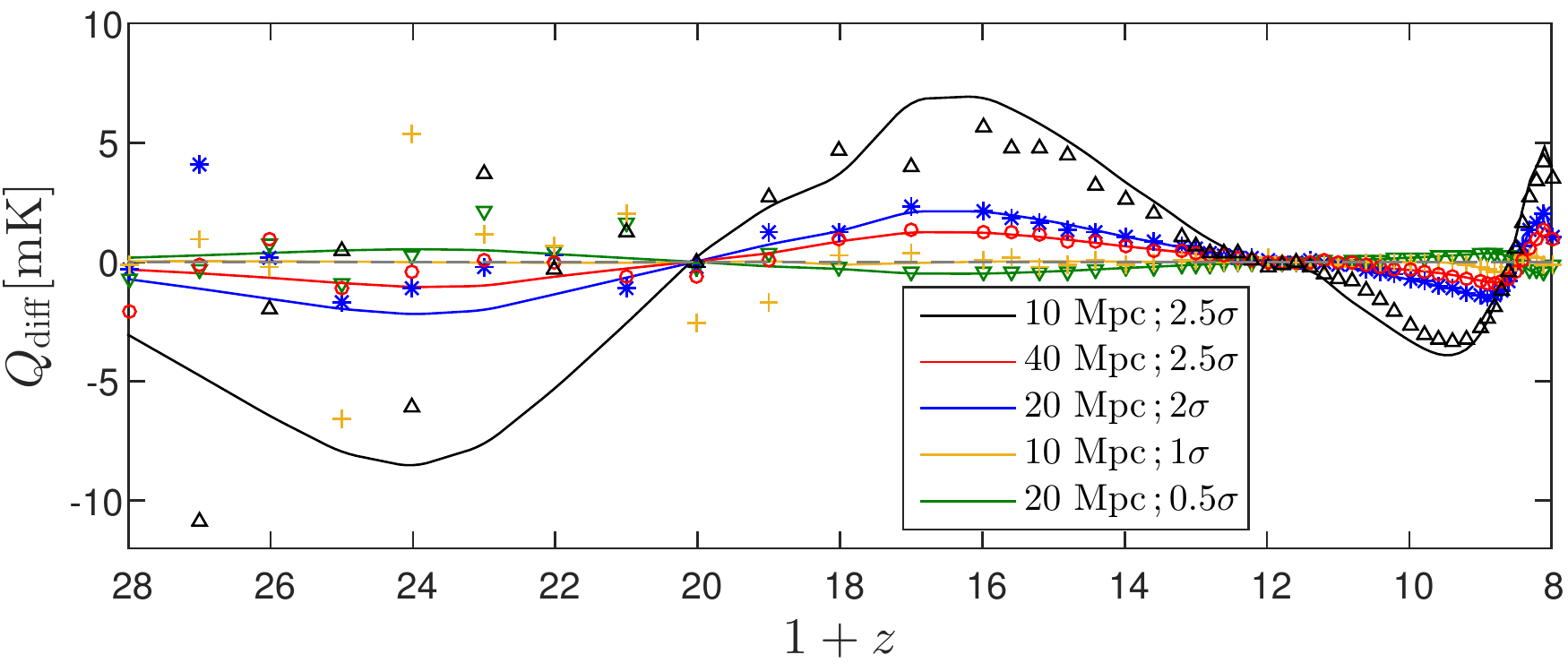}\hspace{0.2in}
\includegraphics[width=3.1in]{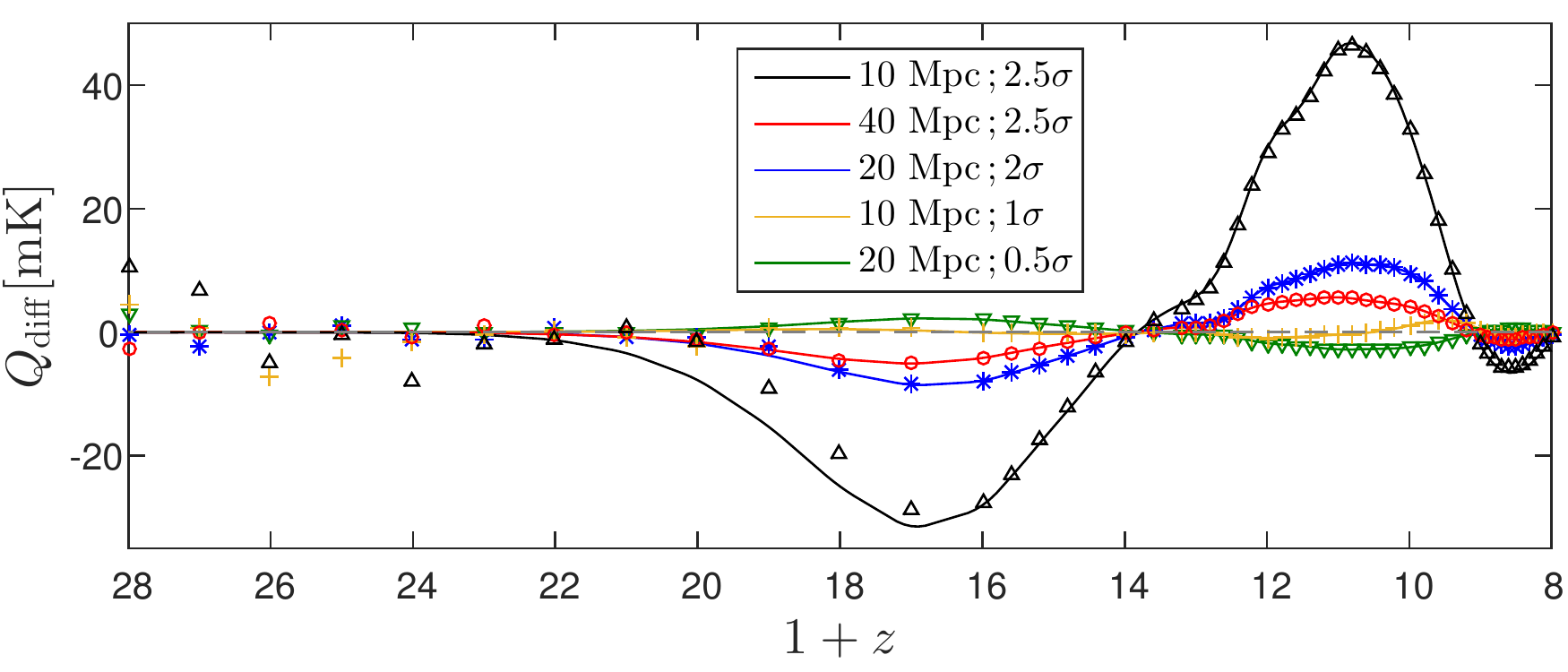}

\includegraphics[width=3.1in]{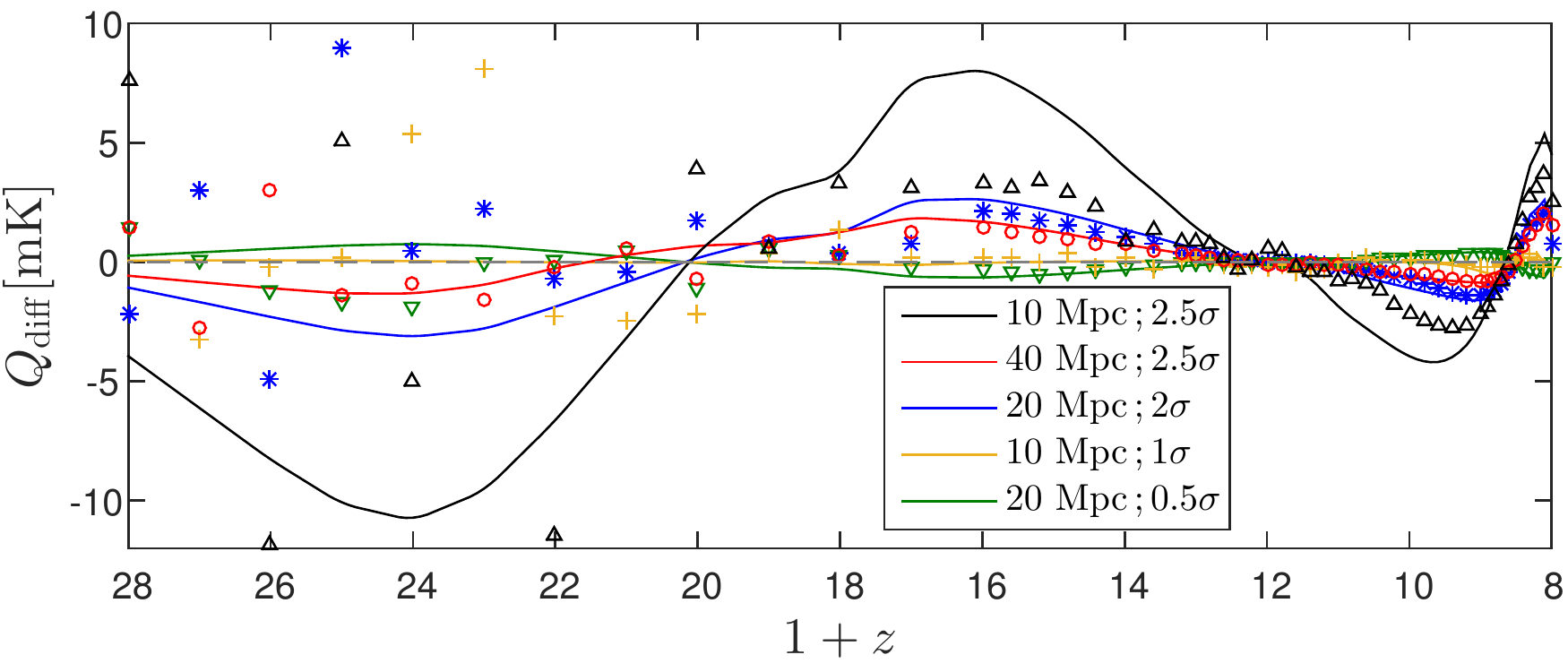}\hspace{0.2in}
\includegraphics[width=3.1in]{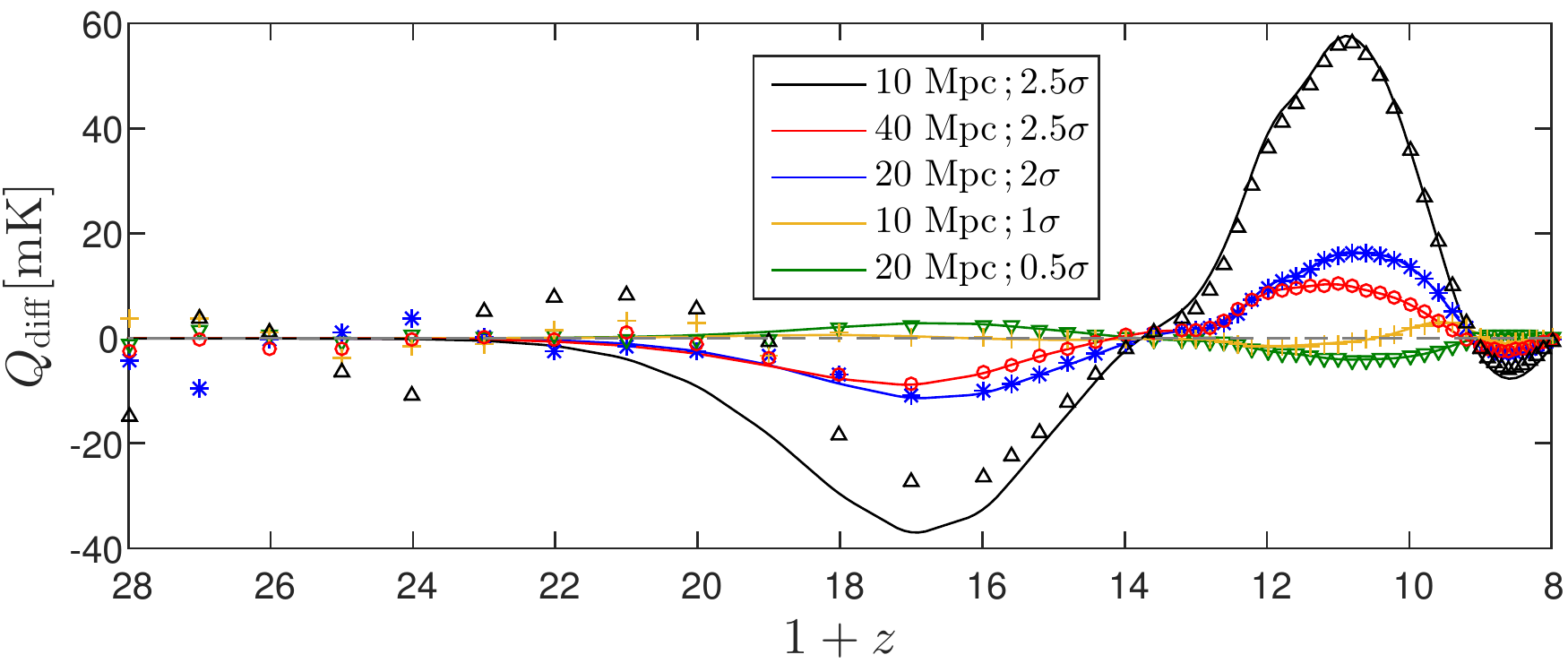}

\caption{Quantile difference (as in Figure~\ref{fig:DifAll}) as a
  function of redshift, for several configurations of threshold $t$
  and smoothing radius $R$. Left panel: Standard model, Right panel:
  Massive model. Top panels use 21-cm images with added 3-D smoothing
  as a first step of the data analysis (our default case throughout
  the paper, see section~\ref{Sec:thermal}), while the bottom panels
  illustrate the same without 3-D smoothing, i.e., using ``raw''
  images with only the inevitable 2-D smoothing that represents the
  observational angular resolution. Symbols in the various panels
  represent the estimated statistic using the map with added noise.}

\label{fig:DifCases}
\end{figure*}

Figure~\ref{fig:DifCases} shows the quantile difference for the
Standard and Massive models, with various choices of threshold $t$ and
comoving radius $R$.  Here the signal can change sign, and is lower in
absolute value than the quantile average shown earlier. Also, the
shape depends more strongly on the choice of $t$ and $R$. In
particular, the low-threshold curves change sign compared to the
high-threshold ones, and for 1$\sigma$ the signal almost
vanishes. This is the reason for us choosing a 2$\sigma$ threshold
(and $R=20\, \rm Mpc$) as our main configuration throughout this
paper. Figure~\ref{fig:DifCases} also shows an example of the results
obtained when we do not add 3-D smoothing at radius $R$ (as discussed
in section~\ref{Sec:thermal}). The results for the statistic measured
from the noise-less 21-cm images (the curves in the figure) are
qualitatively similar to the case with 3-D smoothing, but higher in
absolute value (as there is less smoothing of the 21-cm signal).
However, the reconstructed signal from noisy data is significantly
worse in tracing the correct signal-only result. This shows that 3-D
smoothing removes thermal noise more effectively than it reduces the
21- cm signal, and justifies our inclusion of 3-D smoothing throughout
this work.

\subsection{Normalized quantile average and kurtosis}

\label{sec:kur}

\begin{figure*}
\centering
\includegraphics[width=7in]{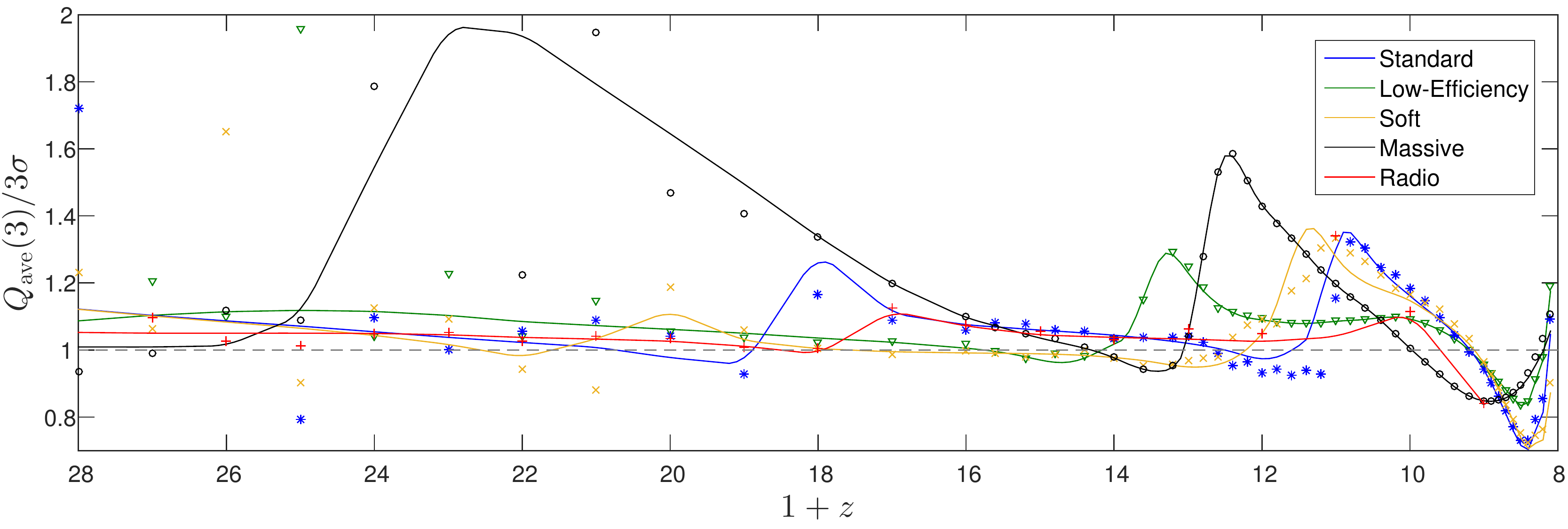}
\includegraphics[width=7in]{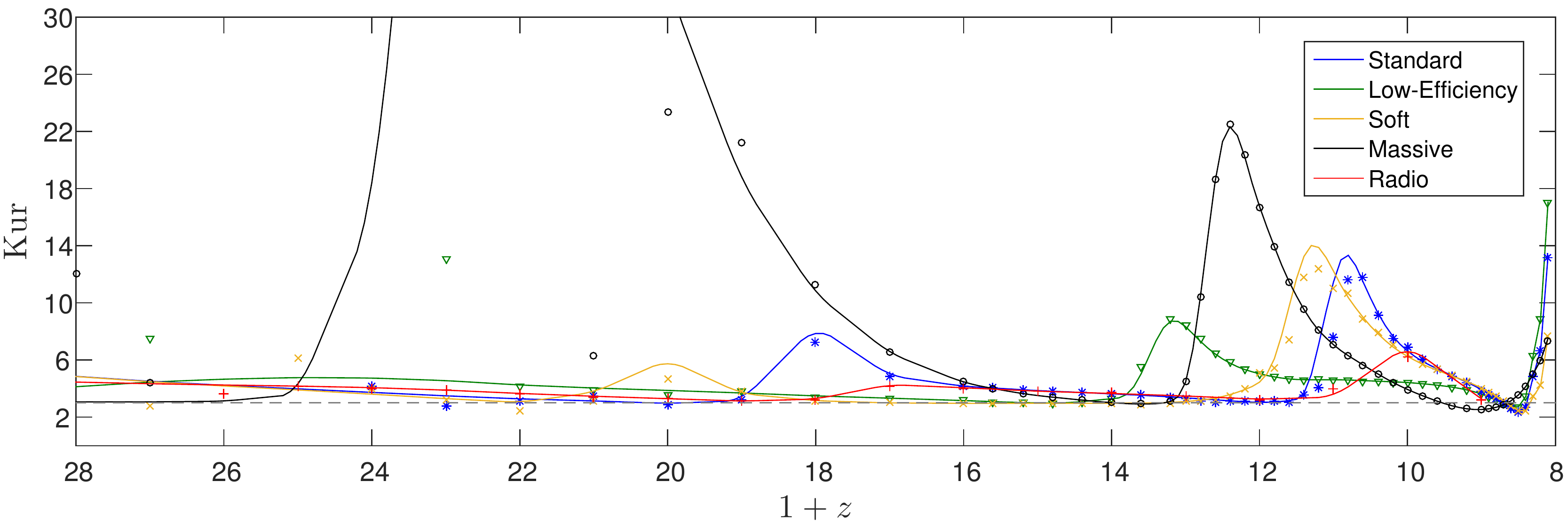}

\caption{Top panel: our alternate kurtosis measure, $Q_{\rm ave}(t)/(t
  \sigma)$ with $t=3$ and $R=20\, \rm Mpc$, shown as a function of
  redshift for our five models. Bottom panel: kurtosis as a function
  of redshift. The symbols in each panel represent the same estimated
  statistic using the map with added noise. The horizontal gray dashed
  line in each panel is the corresponding value for a normal
  distribution. The $y$-axis was truncated in the bottom panel due to
  the fact that the kurtosis goes to infinity as the variance goes to
  zero (near $z=21$).}

\label{fig:Kurtosis}
\end{figure*}

In section~\ref{sec:thresh} we explored the threshold dependence of
the quantile average. Taking the average removes the asymmetry and
with it any sensitivity to the skewness of the distribution. Comparing
the threshold dependence of the quantile average to a Gaussian is thus
most sensitive to the kurtosis. Specifically, we take the normalized
averages from the top panels of Figure~\ref{fig:Threshold} and divide
by $\sigma(z)$. This quantity, $Q_{\rm ave}(t)/(t \sigma)$, which for
a Gaussian would equal unity (independent of $t$), corresponds roughly
to the distribution's kurtosis. The kurtosis is defined as
\begin{equation}
  \rm{Kur}(\textit{X})=\frac{E[(\textit{X}-\mu)^4]}{\sigma^4}\ ,
\end{equation}
and equals 3 for a Gaussian. Figure~\ref{fig:Kurtosis} shows these two
quantities for all five models as a function of redshift with the main
configuration parameters (2$\sigma$ threshold with $R=20\, \rm
Mpc$). We chose $t=3$ because it gave results qualitatively somewhat
more similar to the kurtosis than using $t=2$. The kurtosis estimation
from the noisy map was done using the formula:
\begin{eqnarray}
  \rm{Kur}_{[\rm{est}]}=  \{ \rm{Kur}(\textit{S+N}_1)
  \rm{Var}^2(\textit{S+N}_1)-\rm{Kur}(\textit{N}_2)
  Var^2(\textit{N}_2) \nonumber \\
  - 6[\rm{Var}(\textit{S+N}_1)-\rm{Var}(\textit{N}_2)]
  \rm{Var}(\textit{N}_2) \} \\  \nonumber
   / [\rm{Var}(\textit{S+N}_1)-\rm{Var}(\textit{N}_2)]^2\ ,
\end{eqnarray}
which is easily derived from assuming that the thermal noise is
Gaussian and independent of the signal. As before, here $N_1$ is the
thermal noise added to the signal and $N_2$ is an
independently-generated thermal noise map.

Both the kurtosis and our alternate measure can be measured accurately
from noisy data up to $z \sim 18$. The definitions (which involve
division by $\sigma$) makes the kurtosis (and to a lesser degree the
skewness) especially sensitive to redshifts at which the variance of
the signal is particularly low (i.e., approaches zero, and becomes
difficult to measure accurately). These are the points where the
magnitude of the kurtosis (and of the alternate kurtosis) peaks.
Examples of this can be seen at $z=10$ for the Standard and Soft
models, where the kurtosis estimation deviates from the real
signal-only curve, and at $z > 20$ for the Massive model and $z > 22$
for the Low-Efficiency model. These are redshifts where $\sigma$
approaches zero according to Figure~\ref{fig:AveAll}.\

\section{Summary and Conclusion}

We have suggested quantile-based statistics as a new method for
measuring non-Gaussianities in the 21-cm signal via tomography
maps. This method is complementary to the global signal and power
spectrum which are commonly used and not sensitive to non-Gaussian
aspects such as the asymmetry of the temperature fluctuation
distribution. Quantiles offer a simple, robust and flexible statistic
that is easy to correct for thermal noise. Also, quantiles can be used
to probe the variance, skewness, and kurtosis of the temperature
distribution. The flexibility comes through the ability to choose
different thresholds in the quantile measures. The robustness comes
from being less sensitive to outliers than common statistics that
integrate over the entire distribution function. The simplicity comes
in the noise-correction, which for each quantile measure is done
simply like correcting the variance, i.e., by subtracting the squares
using an independent noise-only map (eq.~\ref{e:est}).

We used mock signals from five possible astrophysical models, covering
the full redshift range of the SKA and exploring a much wider range of
possible signals than previous investigations of non-Gaussian
statistics. This included models with different spectra of the X-ray
heating sources (Soft vs.\ Standard model), different characteristic
masses of galactic halos (Massive vs.\ Standard), different
star-formation and X-ray efficiencies (Low-Efficiency vs.\ Standard),
as well as an exotic model with an excess early radio background
motivated by the EDGES global 21-cm detection. To the single images we
added mock thermal noise according to the expected level for upcoming
observations with the SKA. We tried various smoothing/resolution radii
$R$ of the signal. Varying $R$ allows us to explore various distance
scales, similar to looking at $k$ modes of the power
spectrum. Together with the profile analysis shown in
Figure~\ref{fig:profiles}, this can yield a broad picture of the
spatial behavior of the signal and illuminate the physical processes
involved. For our quantile statistics, we found it advantageous to
add, as an initial analysis step, 3-D smoothing at the same radius
$R$, as this smoothed out the noise more effectively than the signal.

We based our main statistical measures on upper and lower quantiles,
$Q_+$ and $Q_-$, at threshold $t$ defined as containing a cumulative
probability corresponding to a normal distribution, with $t$ in units
of $\sigma$. We then took the symmetric average $Q_{\rm ave}$
(eq.~\ref{e:ave}), which approximately corresponds to measuring the
standard deviation, and the difference $Q_{\rm diff}$
(eq.~\ref{e:diff}), which approximately measures the skewness. We also
showed that the normalized average $Q_{\rm ave}(t)/(t \sigma)$
approximately measures the kurtosis. The threshold dependence of
$Q_{\rm ave}$ (Figure~\ref{fig:Threshold}) can hold more information
that might be explored. For example, we noticed a peak threshold value
in the Radio model (at some redshifts) that does not appear in the
Standard model.

We found that both our statistical measures and the corresponding
standard measures of non-Gaussianity can be measured out to high
redshift with the SKA, often out to $z>20$ and including the redshift
of the Ly$\alpha$ peak. This was the case after accounting for the
expected angular resolution and thermal noise of the SKA (i.e.,
SKA1-Low). This is especially true if the EDGES measurement by
\citet{Bowman:2018} is confirmed, as it implies a stronger amplitude
of 21-cm fluctuations (as exemplified by our Radio model). Generally,
each of our five different astrophysical models has a substantially
different cosmic history, as measured by each statistic (all five
models are shown in Figures~\ref{fig:global}, \ref{fig:profiles},
\ref{fig:AveAll}, \ref{fig:DifAll}, and \ref{fig:Kurtosis}). Thus, the
variation of parameters among the models shows that the minimum
galactic halo mass, the star-formation and X-ray efficiencies, and the
X-ray spectrum, can all be constrained if these statistics are
measured.

With the SKA we will be able to directly image cosmic dawn for the
first time in history. It is necessary to have a variety of methods
and tools that can be applied on the collected data in order to fully
extract the potential it holds. Of course, we have only taken a first
step here, and the next step is to consider more realistic SKA data
with foreground residuals. We expect that the flexibility and
robustness of the quantile statistics will help to deal with that as
well. On the optimistic side, we note that we have used here a
simulation box with volume approximately equal to that of a single SKA
field, while SKA observations will create large surveys covering
multiple fields. Thus, 21-cm cosmology with the SKA holds great
promise.

\section{Acknowledgments}

This project/publication was made possible for AB and RB through the
support of a grant from the John Templeton Foundation. The opinions
expressed in this publication are those of the authors and do not
necessarily reflect the views of the John Templeton Foundation. AB and
RB were also supported by the ISF-NSFC joint research program (grant
No.\ 2580/17). AF was supported by the Royal Society University
Research Fellowship.

\end{document}